\documentclass[
amsmath,amssymb, aps,prl,onecolumn, superscriptaddress, showpacs
]{revtex4-1}
\usepackage{graphicx}
\usepackage{dcolumn}
\usepackage{bm}
\usepackage{mathtools}
\usepackage{xcolor}
\usepackage{float}
\usepackage{amsmath}
\graphicspath{{pics/}}
\usepackage{framed}
\usepackage{qcircuit}
\usepackage{caption}
\usepackage{subfigure}

\def \ket#1{\mathinner{|{#1}\rangle}}

\def \red #1 {\textcolor{red}{#1}}

\usepackage{tikz}
\usepackage{circuitikz}
\usetikzlibrary{backgrounds,fit,decorations.pathreplacing,calc,backgrounds}  

\begin{document}


\title{A quantum circuit design of AES requiring fewer quantum qubits and gate operations}

\author{ZeGuo Wang}
\affiliation{State Key Laboratory of Low-Dimensional Quantum Physics and Department of Physics, Tsinghua University, Beijing 100084, China}

\author{ShiJie Wei}
\email{weishijie@tsinghua.edu.cn}
\affiliation{Beijing Academy of Quantum Information Sciences,  Beijing 100193, China}
\affiliation{State Key Laboratory of Low-Dimensional Quantum Physics and Department of Physics, Tsinghua University, Beijing 100084, China}
\author{GuiLu Long}
\email{gllong@tsinghua.edu.cn}
\affiliation{State Key Laboratory of Low-Dimensional Quantum Physics and Department of Physics, Tsinghua University, Beijing 100084, China}
\affiliation{ Beijing National Research Center for Information Science and Technology and School of Information Tsinghua University, Beijing 100084, China}
\affiliation{Beijing Academy of Quantum Information Sciences,  Beijing 100193, China}
\affiliation{Frontier Science Center for Quantum Information, Beijing 100084, China}

\date{\today}

\begin{abstract}
Advanced Encryption Standard(AES) is one of the most widely used block ciphers nowadays, and  has been established as an encryption standard in 2001. 
Here we design AES-128 and the sample-AES(S-AES) quantum circuits for deciphering. In the quantum circuit of AES-128, we perform an affine transformation for the  SubBytes part to solve the problem that the initial state of the output qubits in SubBytes is not the $\ket{0}^{\otimes 8}$ state. After that, we are able to encode the new round sub-key on the qubits encoding the previous round sub-key, and this improvement  reduces the number of qubits used by 224 compared with Langenberg et al.'s implementation. For S-AES, a complete quantum circuit is presented with only 48 qubits, which is already within the reach of existing noisy intermediate-scale quantum computers.
\end{abstract}

\maketitle


\section{\label{sec1:level1}Introduction}
Classical cryptography are divided into asymmetric cryptography and symmetric cryptography~\cite{bellare2005introduction}. RSA cryptography~\cite{rsa} is a typical asymmetric cipher, whose security relies on our inability to effectively factor large integers. However, RSA cryptography will be broken in the face of quantum computer. Shor's algorithm~\cite{shor} can effectively solve the problem in $O((\log{N})^3)$ time, where N is the large integer to be factorized. AES is a typical symmetric cryptography~\cite{AES}. Grover's quantum algorithm~\cite{grover,long2001grover}, finding a marked item from an unsorted database with square-root speedup, can attack the AES cryptography. Fortunately, the AES is not fully broken. By doubling the length of the key in AES, its security can be maintained. However, it is vital to study the level of influence of quantum computer on the AES.

Akihiroe et al.~\cite{yamamura2000quantum} studied the impact of quantum algorithms on symmetric cryptography. Such as the AES cryptography, which can be decrypted in $O(2^{n/2})$ time using Grover's algorithm, where $n$ is the key length. Kaplen proposed a quantum version~\cite{kaplan2014quantum} of the classical Meet-in-the-Middle attack~\cite{li2016meet} using the Ambainis quantum algorithm~\cite{ambainis2007quantum}. The time complexity and qubits complexity of Kaplen's algorithm both are $O(2^{2n/3})$. In 2014, Roetteler et al. proposed the quantum related key attack~\cite{roetteler2015note} by using Simon's algorithm~\cite{365701}. Grassl et al.~\cite{2015Applying} presented a quantum implementation of AES in 2015 and analyzed the consumption of quantum resources. This is the first time that the "zig-zag" method is used to reduce the number of qubits. In the "zig-zag" method, the qubits that are used to encode the cipher in the front rounds of encryption can be released and reused to encode later rounds of cipher. Later on, Kim et al.~\cite{kim2018time} made an improvement on SubBytes operation, one of the four operations in AES, in Grassl et al.'s  work, which saves one multiplication operation. In 2018, Almazrooie et al.~\cite{almazrooie2018quantum} proposed a quantum circuit that uses Grover's algorithm to attack S-AES cryptography. Owing to the small size, this is the most likely case to implement in the near-term quantum computing systems. Recently, Langenberg et al. designed a quantum circuit \cite{langenberg2020reducing} for SubBytes based on Boyar et al.'s classical algorithm~\cite{boyar2010new}. It reduced the number of Toffoli gates by $88\%$ compared to Grassl et al.'s  work. 

In this paper, we design an improved key expansion quantum circuit. The SubBytes quantum circuit in~\cite{langenberg2020reducing} is modified firstly, which eliminates the effect that the output qubits in SubBytes are not in $\ket{0}^{\otimes 8}$. Then, we construct the quantum circuit to encode the new round sub-key on the qubits encoding previous round sub-key. This can reduce the cost of qubits by 224 compared to  Langenberg et al.'s work~\cite{langenberg2020reducing}.  Employing the lowest possible number of qubits,  we optimize the quantum circuit of S-AES using the methods used in AES-128. Compared with Almazrooie et al.'s work~\cite{almazrooie2018quantum}, it saves $71\%$ of Toffoli gates, over $66\%$ C-NOT gates and one third(24) for qubits. 

This paper is organized as follows: the first part briefly reviews the structure of AES algorithm and the quantum circuit. Then an improved key expansion quantum circuit is proposed, and the SubBytes quantum circuit from Ref~\cite{langenberg2020reducing} is modified. The quantum circuit of S-AES is optimized using the same methods as in AES. Next, we analyze the complexity of AES-128 and S-AES quantum circuits  and compared with the results in Ref~\cite{langenberg2020reducing} and Ref~\cite{almazrooie2018quantum} respectively. Finally, discussion and conclusion are presented.


\section{\label{sec1:level1}Introduction to AES-128}
\begin{figure*}[htb]
\center{\includegraphics[width=15cm]{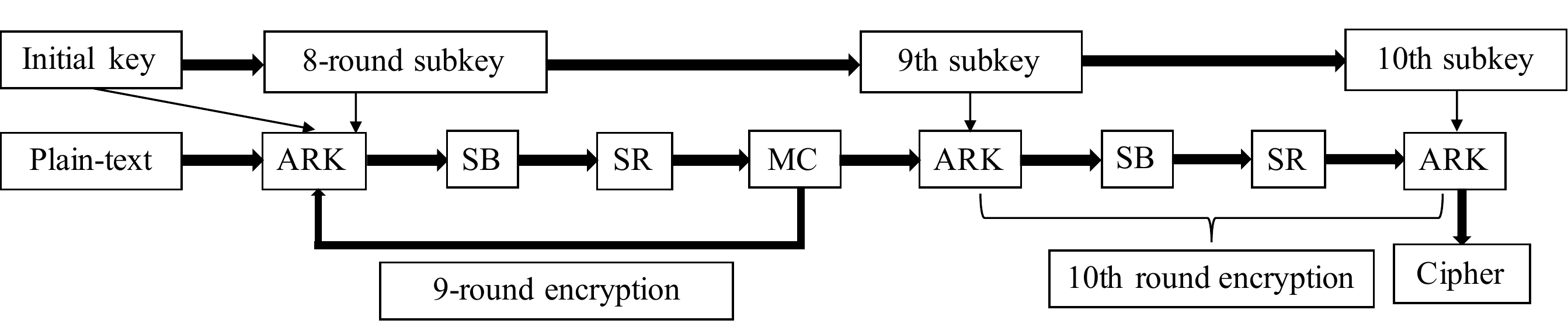}}
\caption{AES-128: ARK represents AddRoundKey, SB represents SubBytes, SR represents ShiftRows, and  MC represents MixColumns.}
\label{AES-128}
\end{figure*}

As a symmetric block cipher, AES-128 uses the same 128-bit key each time to encrypt 128-bit plain-text. Both the key and the plain-text are arranged into a rectangular array of $4\times4$ bytes. The entire encryption process contains 4 encryption operations, namely AddRoundKey, SubBytes, ShiftRows and MixColumns. In the first nine rounds of encryption, the above 4 encryption operations are executed in order, and the tenth round of encryption contains two AddRoundKey operations but without MixColumns operation. Therefore, in addition to the 128-bit initial key, 10 rounds sub-key are needed to be generated for AddRoundKey. The specific flow chart is shown in Figure \ref{AES-128}. In the key expansion operation, transposition and SubBytes are mainly used. Below we briefly introduce the 4 kinds of encryption operations.

AddRoundKey implements the bit-wise XOR of the round keys, and it needs 128 C-NOT gates each time and can be executed in parallel.

ShiftRows is a particular permutation of the current AES state. For the first row in the plain-text array of $4\times4$ bytes, we do nothing; for the second row, rotate one unit(a byte) to the left; for the third row, rotate two units to the left; for the last row, rotate three units to the left. All of the above operations are equivalent to adjust the position of the bytes, so no additional operations are required.

MixColumns operates on an entire column of the plain-text array at a time as an affine transformation, so MixColumns is equivalent to a $32\times32$ matrix acting on every column. A MixColumns requires 277 C-NOT gates.

SubBytes is the most important and complex part of the entire algorithm which prevents linear correspondence between the plain-text and the cipher-text. The main function of SubBytes is to map one byte to another through the S-box. SubBytes can be achieved by performing the inverse operation for each byte in the $4\times4$ bytes array on the finite field, and then performing an affine transformation for every byte.

Next, we briefly introduce the key expansion algorithm.
The 16-byte initial key is arranged into a $4\times4$ array, which is divided into $k_0,k_1,k_2,k_3$ four parts, each part has 4 bytes. Firstly, for the 4 bytes of $k_3$, we loop them to the left by one unit(a byte) and do SubBytes operation for each of the byte, followed by performing XOR operations with a known 32-bit string. Finally, XOR operations between $k_0$ and the above results are performed to get $k_4$. $k_5$ can be obtained by making XOR operations between $k_1$ and $k_4$. And so on, $k_6$, $k_7$ can be gained. A round key expansion is finished. The last 9-round sub-key can be gained by repeating the above process. The 32 known bits that perform XOR operations with the 4-byte key are different in each round, a total of 16 Pauli-X gates are required for 10 rounds.
In Langenberg et al.'s work, the key expansion quantum circuit needs 224 qubits to generate the 10-round sub-key. 

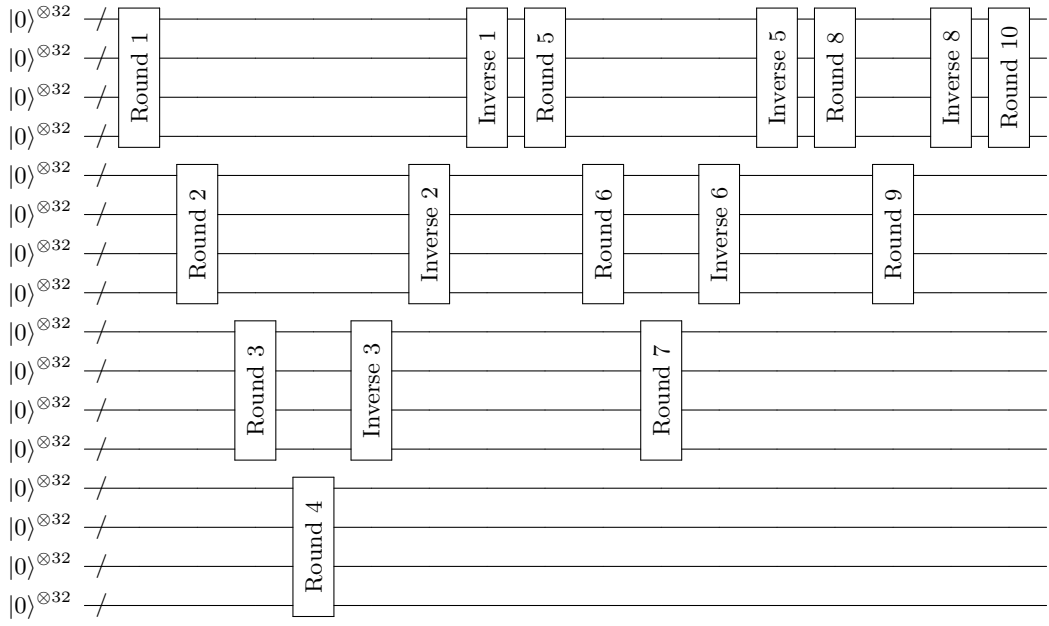
\begin{figure*}
\centerline{
\Qcircuit @C=.7em @R=.7em {
\lstick{\ket{0}^{\otimes 32}} &{/} \qw &\multigate{3}{\rotatebox{90}{Round 1}} &\qw &\qw &\qw &\qw &\qw &\multigate{3}{\rotatebox{90}{Inverse 1}} &\multigate{3}{\rotatebox{90}{Round 5}} &\qw &\qw &\qw &\multigate{3}{\rotatebox{90}{Inverse 5}} &\multigate{3}{\rotatebox{90}{Round 8}} &\qw &\multigate{3}{\rotatebox{90}{Inverse 8}} &\multigate{3}{\rotatebox{90}{Round 10}} &\qw \\
\lstick{\ket{0}^{\otimes 32}} &{/} \qw &\ghost{\rotatebox{90}{Round 1}} &\qw &\qw &\qw &\qw &\qw &\ghost{\rotatebox{90}{Inverse 1}} &\ghost{\rotatebox{90}{Round 5}} &\qw &\qw &\qw &\ghost{\rotatebox{90}{Inverse 5}} &\ghost{\rotatebox{90}{Round 8}} &\qw &\ghost{\rotatebox{90}{Inverse 8}} &\ghost{\rotatebox{90}{Round 10}} &\qw \\
\lstick{\ket{0}^{\otimes 32}} &{/} \qw &\ghost{\rotatebox{90}{Round 1}} &\qw &\qw &\qw &\qw &\qw &\ghost{\rotatebox{90}{Inverse 1}} &\ghost{\rotatebox{90}{Round 5}} &\qw &\qw &\qw &\ghost{\rotatebox{90}{Inverse 5}} &\ghost{\rotatebox{90}{Round 8}} &\qw &\ghost{\rotatebox{90}{Inverse 8}} &\ghost{\rotatebox{90}{Round 10}} &\qw \\
\lstick{\ket{0}^{\otimes 32}} &{/} \qw &\ghost{\rotatebox{90}{Round 1}} &\qw &\qw &\qw &\qw &\qw &\ghost{\rotatebox{90}{Inverse 1}} &\ghost{\rotatebox{90}{Round 5}} &\qw &\qw &\qw &\ghost{\rotatebox{90}{Inverse 5}} &\ghost{\rotatebox{90}{Round 8}} &\qw &\ghost{\rotatebox{90}{Inverse 8}} &\ghost{\rotatebox{90}{Round 10}} &\qw \\
\lstick{\ket{0}^{\otimes 32}} &{/} \qw &\qw &\multigate{3}{\rotatebox{90}{Round 2}} &\qw &\qw &\qw &\multigate{3}{\rotatebox{90}{Inverse 2}} &\qw &\qw &\multigate{3}{\rotatebox{90}{Round 6}} &\qw &\multigate{3}{\rotatebox{90}{Inverse 6}} &\qw &\qw &\multigate{3}{\rotatebox{90}{Round 9}} &\qw &\qw &\qw \\
\lstick{\ket{0}^{\otimes 32}} &{/} \qw &\qw &\ghost{\rotatebox{90}{Round 2}} &\qw &\qw &\qw &\ghost{\rotatebox{90}{Inverse 2}} &\qw &\qw &\ghost{\rotatebox{90}{Round 6}} &\qw &\ghost{\rotatebox{90}{Inverse 6}} &\qw &\qw &\ghost{\rotatebox{90}{Round 9}} &\qw &\qw &\qw \\
\lstick{\ket{0}^{\otimes 32}} &{/} \qw &\qw &\ghost{\rotatebox{90}{Round 2}} &\qw &\qw &\qw &\ghost{\rotatebox{90}{Inverse 2}} &\qw &\qw &\ghost{\rotatebox{90}{Round 6}} &\qw &\ghost{\rotatebox{90}{Inverse 6}} &\qw &\qw &\ghost{\rotatebox{90}{Round 9}} &\qw &\qw &\qw \\
\lstick{\ket{0}^{\otimes 32}} &{/} \qw &\qw &\ghost{\rotatebox{90}{Round 2}} &\qw &\qw &\qw &\ghost{\rotatebox{90}{Inverse 2}} &\qw &\qw &\ghost{\rotatebox{90}{Round 6}} &\qw &\ghost{\rotatebox{90}{Inverse 6}} &\qw &\qw &\ghost{\rotatebox{90}{Round 9}} &\qw &\qw &\qw \\
\lstick{\ket{0}^{\otimes 32}} &{/} \qw &\qw &\qw &\multigate{3}{\rotatebox{90}{Round 3}} &\qw &\multigate{3}{\rotatebox{90}{Inverse 3}} &\qw &\qw &\qw &\qw &\multigate{3}{\rotatebox{90}{Round 7}} &\qw &\qw &\qw &\qw &\qw &\qw &\qw \\
\lstick{\ket{0}^{\otimes 32}} &{/} \qw &\qw &\qw &\ghost{\rotatebox{90}{Round 3}} &\qw &\ghost{\rotatebox{90}{Inverse 3}} &\qw &\qw &\qw &\qw &\ghost{\rotatebox{90}{Round 7}} &\qw &\qw &\qw &\qw &\qw &\qw &\qw \\
\lstick{\ket{0}^{\otimes 32}} &{/} \qw &\qw &\qw &\ghost{\rotatebox{90}{Round 3}} &\qw &\ghost{\rotatebox{90}{Inverse 3}} &\qw &\qw &\qw &\qw &\ghost{\rotatebox{90}{Round 7}} &\qw &\qw &\qw &\qw &\qw &\qw &\qw \\
\lstick{\ket{0}^{\otimes 32}} &{/} \qw &\qw &\qw &\ghost{\rotatebox{90}{Round 3}} &\qw &\ghost{\rotatebox{90}{Inverse 3}} &\qw &\qw &\qw &\qw &\ghost{\rotatebox{90}{Round 7}} &\qw &\qw &\qw &\qw &\qw &\qw &\qw \\
\lstick{\ket{0}^{\otimes 32}} &{/} \qw &\qw &\qw &\qw &\multigate{3}{\rotatebox{90}{Round 4}} &\qw &\qw &\qw &\qw &\qw &\qw &\qw &\qw &\qw &\qw &\qw &\qw &\qw \\
\lstick{\ket{0}^{\otimes 32}} &{/} \qw &\qw &\qw &\qw &\ghost{\rotatebox{90}{Round 4}} &\qw &\qw &\qw &\qw &\qw &\qw &\qw &\qw &\qw &\qw &\qw &\qw &\qw \\
\lstick{\ket{0}^{\otimes 32}} &{/} \qw &\qw &\qw &\qw &\ghost{\rotatebox{90}{Round 4}} &\qw &\qw &\qw &\qw &\qw &\qw &\qw &\qw &\qw &\qw &\qw &\qw &\qw \\
\lstick{\ket{0}^{\otimes 32}} &{/} \qw &\qw &\qw &\qw &\ghost{\rotatebox{90}{Round 4}} &\qw &\qw &\qw &\qw &\qw &\qw &\qw &\qw &\qw &\qw &\qw &\qw &\qw \\
}
}
\caption{Quantum circuit of AES-128: a line represents 32 qubits}
\label{QCA}
\end{figure*}

In the last part of this section, the quantum circuit of AES-128 is shown in Figure \ref{QCA}. Round $n$($n=1,2...10$) represents the encryption round and Inverse $n$ represents the inverse operation for the corresponding encryption round. The plain-text is encoded on the qubits that encodes the initial key, which can save 128 qubits used to encode plain-text. We denote the XOR operations between the plain-text and the initial key as $P$.

\section{improvements in quantum implementation of AES-128}
\begin{figure*}
\centerline{
\Qcircuit @C=1em @R=.7em {
\lstick{\ket{0}^{\otimes 8}} &{/} \qw &\multigate{15}{H^{\otimes 128}} &\gate{SB^*} &\qw &\qw &\qw &\multigate{3}{RC} &\ctrl{4} &\qw &\qw &\qw &\qw &\qw &\qw &\qw &\qw &\qw &\qw &\qw &\qw \\
\lstick{\ket{0}^{\otimes 8}} &{/} \qw &\ghost{H^{\otimes 128}} &\qw &\gate{SB^*}  &\qw &\qw &\ghost{RC} &\qw &\ctrl{4} &\qw &\qw &\qw &\qw &\qw &\qw &\qw &\qw &\qw &\qw &\qw \\
\lstick{\ket{0}^{\otimes 8}} &{/} \qw &\ghost{H^{\otimes 128}} &\qw &\qw &\gate{SB^*} &\qw &\ghost{RC} &\qw &\qw &\ctrl{4} &\qw &\qw &\qw &\qw &\qw &\qw &\qw &\qw &\qw &\qw \\
\lstick{\ket{0}^{\otimes 8}} &{/} \qw &\ghost{H^{\otimes 128}} &\qw &\qw &\qw &\gate{SB^*} &\ghost{RC} &\qw &\qw &\qw &\ctrl{4} &\qw &\qw &\qw &\qw &\qw &\qw &\qw &\qw &\qw \\
\lstick{\ket{0}^{\otimes 8}} &{/} \qw &\ghost{H^{\otimes 128}} &\qw &\qw &\qw &\qw &\qw &\targ &\qw &\qw &\qw &\ctrl{4} &\qw &\qw &\qw &\qw &\qw &\qw &\qw &\qw  \\
\lstick{\ket{0}^{\otimes 8}} &{/} \qw &\ghost{H^{\otimes 128}} &\qw &\qw &\qw &\qw &\qw &\qw &\targ &\qw &\qw &\qw &\ctrl{4} &\qw &\qw &\qw &\qw &\qw &\qw &\qw   \\
\lstick{\ket{0}^{\otimes 8}} &{/} \qw &\ghost{H^{\otimes 128}} &\qw &\qw &\qw &\qw &\qw &\qw &\qw &\targ &\qw &\qw &\qw &\ctrl{4} &\qw &\qw &\qw &\qw &\qw &\qw  \\
\lstick{\ket{0}^{\otimes 8}} &{/} \qw &\ghost{H^{\otimes 128}} &\qw &\qw &\qw &\qw &\qw &\qw &\qw &\qw &\targ &\qw &\qw &\qw &\ctrl{4} &\qw &\qw &\qw &\qw &\qw   \\
\lstick{\ket{0}^{\otimes 8}} &{/} \qw &\ghost{H^{\otimes 128}} &\qw &\qw &\qw &\qw &\qw &\qw &\qw &\qw &\qw &\targ &\qw &\qw &\qw &\ctrl{4} &\qw &\qw &\qw &\qw    \\
\lstick{\ket{0}^{\otimes 8}} &{/} \qw &\ghost{H^{\otimes 128}} &\qw &\qw &\qw &\qw &\qw &\qw &\qw &\qw &\qw &\qw &\targ &\qw &\qw &\qw &\ctrl{4} &\qw &\qw &\qw    \\
\lstick{\ket{0}^{\otimes 8}} &{/} \qw &\ghost{H^{\otimes 128}} &\qw &\qw &\qw &\qw &\qw &\qw &\qw &\qw &\qw &\qw &\qw &\targ &\qw &\qw &\qw &\ctrl{4} &\qw &\qw    \\
\lstick{\ket{0}^{\otimes 8}} &{/} \qw &\ghost{H^{\otimes 128}} &\qw &\qw &\qw &\qw &\qw &\qw &\qw &\qw &\qw &\qw &\qw &\qw &\targ &\qw &\qw &\qw &\ctrl{4} &\qw     \\
\lstick{\ket{0}^{\otimes 8}} &{/} \qw &\ghost{H^{\otimes 128}} &\qw &\qw &\qw &\ctrl{-9} &\qw &\qw &\qw &\qw &\qw &\qw &\qw &\qw &\qw &\targ &\qw &\qw &\qw &\qw    \\
\lstick{\ket{0}^{\otimes 8}} &{/} \qw &\ghost{H^{\otimes 128}} &\ctrl{-13} &\qw &\qw &\qw &\qw &\qw &\qw &\qw &\qw &\qw &\qw &\qw &\qw &\qw &\targ &\qw &\qw &\qw    \\
\lstick{\ket{0}^{\otimes 8}} &{/} \qw &\ghost{H^{\otimes 128}} &\qw &\ctrl{-13} &\qw &\qw &\qw &\qw &\qw &\qw &\qw &\qw &\qw &\qw &\qw &\qw &\qw &\targ &\qw &\qw    \\
\lstick{\ket{0}^{\otimes 8}} &{/} \qw &\ghost{H^{\otimes 128}} &\qw &\qw &\ctrl{-13} &\qw &\qw &\qw &\qw &\qw &\qw &\qw &\qw &\qw &\qw &\qw &\qw &\qw &\targ &\qw    \\
}
}
\caption{Key Expansion: $SB^*$ represents the modified SubBytes, RC represents making XOR operations with the known 32-bits string.}
\label{SG1}
\end{figure*}
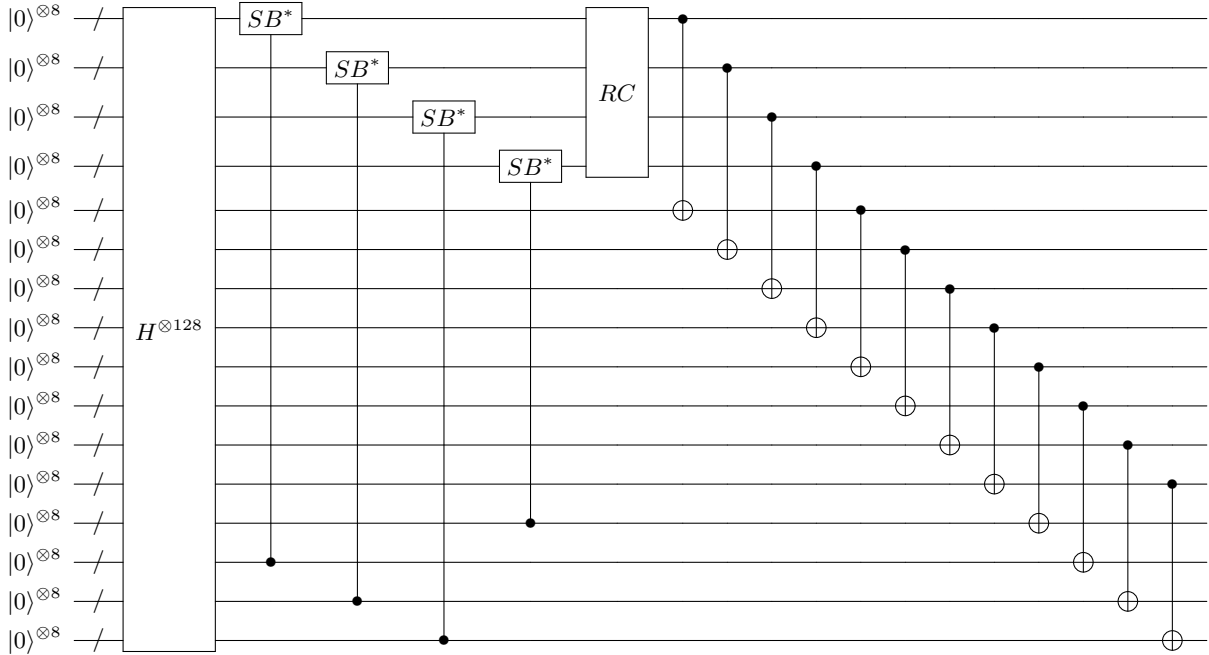

We propose an improved design for key expansion, which does not need the extra qubits for saving the sub-key. It is shown in Figure \ref{SG1}.  

\begin{figure}[H]
\centerline{
\Qcircuit @C=1em @R=.7em {
\lstick{\ket{a_1}} &\qw &\qw &\qw &\ctrl{6} &\qw &\qw &\ctrl{1} &\qw \\
\lstick{\ket{a_2}} &\qw &\qw &\ctrl{2} &\qw &\targ &\ctrl{3} &\targ &\qw \\
\lstick{\ket{a_3}} &\ctrl{3} &\targ &\qw &\qw &\qw &\qw &\qw &\qw \\
\lstick{\ket{a_4}} &\qw &\qw &\targ &\qw &\qw &\qw &\qw &\qw \\
\lstick{\ket{a_5}} &\qw &\qw &\qw &\qw &\qw &\targ &\qw &\qw \\
\lstick{\ket{a_6}} &\targ &\qw &\qw &\qw &\qw &\qw &\qw &\qw \\
\lstick{\ket{a_7}} &\qw &\qw &\ctrl{1} &\targ &\qw &\qw &\qw &\qw \\
\lstick{\ket{a_8}} &\qw &\ctrl{-5} &\targ &\qw &\ctrl{-6} &\qw &\qw &\qw
}
}
\caption{The affine transformation on the output qubits}
\label{ATO}
\end{figure}
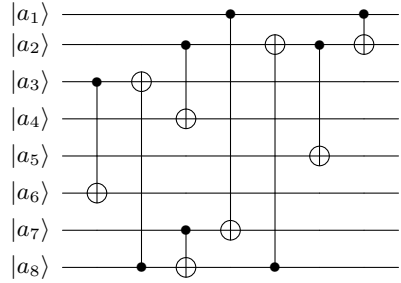

The new design saves 224 qubits because we generate the sub-key in the qubits that encode the previous key, instead of using 32 qubits in $\ket{0}$ state per round to save the new sub-key. The quantum circuit for SubBytes in \cite{langenberg2020reducing}, requiring 16 auxiliary qubits, 55 Toffoli gates, 314 C-NOT gates and 4 NOT gates, can be used in the 10-round encryption but cannot be applied in the key expansion since the initial state of the output qubits in SubBytes are not $\ket{0}^{\otimes 8}$. In order to use this quantum circuit of SubBytes, an affine transformation shown in Figure \ref{ATO} operates on the output qubits which eliminates the effect that the output qubits are not in $\ket{0}^{\otimes 8}$ state, and the final result is the same as the output of SubBytes and the initial state of the output qubits to do XOR operations. It is denoted as $SB^*$. Suppose the input is $X$ and the initial state of output qubits is $Y$ in this step, this process can be written as a function:
\begin{equation}
SB^*(X)=Y \oplus SB(X)
\end{equation}

This affine transformation contains 8 C-NOT gates but does not increase the depth of the entire quantum circuit of SubBytes. So the overall cost of SubBytes in key expansion is 16 auxiliary qubits, 55 Toffoli gates, 322 C-NOT gates and 4 NOT gates.

\section{An improved quantum implementation of S-AES}
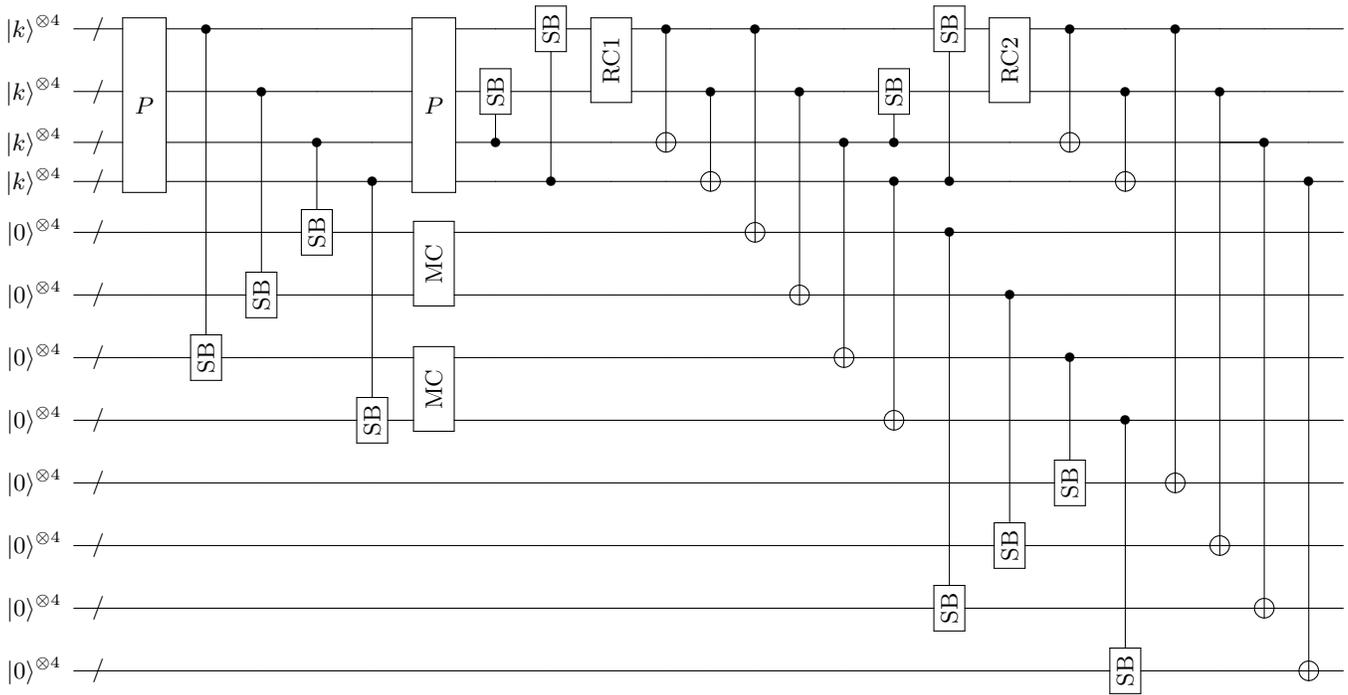
\begin{figure*}[htb]
\centerline{
\Qcircuit @C=1em @R=.7em {
\lstick{\ket{k}^{\otimes 4}} &{/} \qw &\multigate{3}{P} &\ctrl{6} &\qw &\qw &\qw &\multigate{3}{P} &\qw &\gate{\rotatebox{90}{SB}} &\multigate{1}{\rotatebox{90}{RC1}} &\ctrl{2} &\qw &\ctrl{4} &\qw &\qw &\qw &\gate{\rotatebox{90}{SB}} &\multigate{1}{\rotatebox{90}{RC2}} &\ctrl{2} &\qw &\ctrl{8} &\qw &\qw &\qw &\qw \\
\lstick{\ket{k}^{\otimes 4}} &{/} \qw &\ghost{P} &\qw &\ctrl{4} &\qw &\qw &\ghost{P} &\gate{\rotatebox{90}{SB}} &\qw &\ghost{\rotatebox{90}{RC1}} &\qw &\ctrl{2} &\qw &\ctrl{4} &\qw &\gate{\rotatebox{90}{SB}} &\qw &\ghost{\rotatebox{90}{RC2}} &\qw &\ctrl{2} &\qw &\ctrl{8} &\qw &\qw &\qw\\
\lstick{\ket{k}^{\otimes 4}} &{/} \qw &\ghost{P} &\qw &\qw &\ctrl{2} &\qw &\ghost{P} &\ctrl{-1} &\qw &\qw &\targ &\qw &\qw &\qw &\ctrl{4} &\ctrl{-1} &\qw &\qw &\targ &\qw &\qw &\qw &\ctrl{8} \qw &\qw &\qw\\
\lstick{\ket{k}^{\otimes 4}} &{/} \qw &\ghost{P} &\qw &\qw &\qw &\ctrl{4} &\ghost{P} &\qw &\ctrl{-3} &\qw &\qw &\targ &\qw &\qw &\qw &\ctrl{4} &\ctrl{-3} &\qw &\qw &\targ &\qw &\qw &\qw &\ctrl{8} &\qw\\
\lstick{\ket{0}^{\otimes 4}} &{/} \qw &\qw &\qw &\qw &\gate{\rotatebox{90}{SB}} &\qw &\multigate{1}{\rotatebox{90}{MC}} &\qw &\qw &\qw &\qw &\qw &\targ &\qw &\qw &\qw &\ctrl{6} &\qw &\qw &\qw &\qw &\qw &\qw &\qw &\qw\\
\lstick{\ket{0}^{\otimes 4}} &{/} \qw &\qw &\qw &\gate{\rotatebox{90}{SB}} &\qw &\qw &\ghost{\rotatebox{90}{MC}} &\qw &\qw &\qw &\qw &\qw &\qw &\targ &\qw &\qw &\qw &\ctrl{4} &\qw &\qw &\qw &\qw &\qw &\qw  &\qw\\
\lstick{\ket{0}^{\otimes 4}} &{/} \qw &\qw &\gate{\rotatebox{90}{SB}} &\qw &\qw &\qw &\multigate{1}{\rotatebox{90}{MC}} &\qw &\qw &\qw &\qw &\qw &\qw &\qw &\targ &\qw &\qw &\qw &\ctrl{2} &\qw &\qw &\qw &\qw &\qw &\qw\\
\lstick{\ket{0}^{\otimes 4}} &{/} \qw &\qw &\qw &\qw &\qw &\gate{\rotatebox{90}{SB}} &\ghost{\rotatebox{90}{MC}} &\qw &\qw &\qw &\qw &\qw &\qw &\qw &\qw &\targ &\qw &\qw &\qw &\ctrl{4} &\qw &\qw &\qw &\qw &\qw \\
\lstick{\ket{0}^{\otimes 4}} &{/} \qw &\qw &\qw &\qw &\qw &\qw &\qw &\qw &\qw &\qw &\qw &\qw &\qw &\qw &\qw &\qw &\qw &\qw &\gate{\rotatebox{90}{SB}} &\qw &\targ &\qw &\qw &\qw &\qw \\
\lstick{\ket{0}^{\otimes 4}} &{/} \qw &\qw &\qw &\qw &\qw &\qw &\qw &\qw &\qw &\qw &\qw &\qw &\qw &\qw &\qw &\qw &\qw &\gate{\rotatebox{90}{SB}} &\qw &\qw &\qw &\targ &\qw &\qw &\qw \\
\lstick{\ket{0}^{\otimes 4}} &{/} \qw &\qw &\qw &\qw &\qw &\qw &\qw &\qw &\qw &\qw &\qw &\qw &\qw &\qw &\qw &\qw &\gate{\rotatebox{90}{SB}} &\qw &\qw &\qw &\qw &\qw &\targ &\qw &\qw\\
\lstick{\ket{0}^{\otimes 4}} &{/} \qw &\qw &\qw &\qw &\qw &\qw &\qw &\qw &\qw &\qw &\qw &\qw &\qw &\qw &\qw &\qw &\qw &\qw &\qw &\gate{\rotatebox{90}{SB}} &\qw &\qw &\qw &\targ &\qw
}
}
\caption{An more economic quantum circuit for S-AES: P represents the XOR operations between the key and the plain-text; SB represents the SubBytes; MC represents the MixColumns; RC represents XOR operations with the known 8-bit string; the last 16-qubit represents the cipher}
\label{QCS}
\end{figure*}

Here, the same method is used in S-AES. It saves 16 qubits without increasing the gate complexity. Figure \ref{QCS} shows the quantum circuit of S-AES.

For S-AES, we design a SubBytes quantum circuit by using the Boolean functions corresponding to the inverse in finite field $GF(2^4)$. The inverse representation in $GF(2^4)$ based on the normal basis is as follows :
\begin{equation}
\begin{array}{l}
y_{1}=x_{2} x_{3} x_{4}+x_{1} x_{3}+x_{2} x_{3}+x_{3}+x_{4} \\
y_{2}=x_{1} x_{3} x_{4}+x_{1} x_{3}+x_{2} x_{3}+x_{2} x_{4}+x_{4} \\
y_{3}=x_{1} x_{2} x_{4}+x_{1} x_{3}+x_{1} x_{4}+x_{1}+x_{2} \\
y_{4}=x_{1} x_{2} x_{3}+x_{1} x_{3}+x_{1} x_{4}+x_{2} x_{4}+x_{2}
\end{array}
\label{BF}
\end{equation}
Where $x_{i}$ represents the input qubit and $y_{i}$ represents the output qubit$(i=1,2,3,4)$. There are 4 mapping matrices between this normal basis and polynomial basis, the simplest of which is as follow:
\begin{equation}
\left[\begin{array}{llll}
0 & 1 & 1 & 1 \\
0 & 1 & 0 & 1 \\
1 & 0 & 0 & 1 \\
0 & 0 & 1 & 1
\end{array}\right]
\end{equation}

\begin{figure}
\centering
\subfigure[Polynomial basis to normal basis]{
\Qcircuit @C=1em @R=.7em {
\lstick{\ket{a_3}} &\targ &\qw &\qw &\qw &\qw \\
\lstick{\ket{a_2}} &\qw &\targ &\qw &\ctrl{1} &\qw \\
\lstick{\ket{a_1}} &\qw &\qw &\ctrl{1} &\targ &\qw \\
\lstick{\ket{a_0}} &\ctrl{-3} &\ctrl{-2} &\targ &\qw &\qw
}
}
\hspace{1cm}
\subfigure[Normal basis to polynomial basis]{
\Qcircuit @C=1em @R=.7em {
\lstick{\ket{x_1}} &\targ &\ctrl{3} &\qw &\qw &\qw \\
\lstick{\ket{x_2}} &\ctrl{-1} &\qw &\targ &\qw &\qw \\
\lstick{\ket{x_3}} &\qw &\qw &\qw &\targ &\qw\\
\lstick{\ket{x_4}} &\qw &\targ &\ctrl{-2} &\ctrl{-1} &\qw
}
}
\caption{Basis transformation}
\label{TS}
\end{figure}
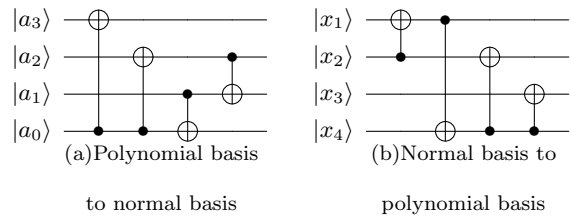

This matrix transforms the polynomial basis to the normal basis, 
and the inverse matrix transforms the normal basis to the polynomial basis. Figure \ref{TS} is the corresponding quantum circuit. It requires 4 C-NOT gates to implement the basis transformation.

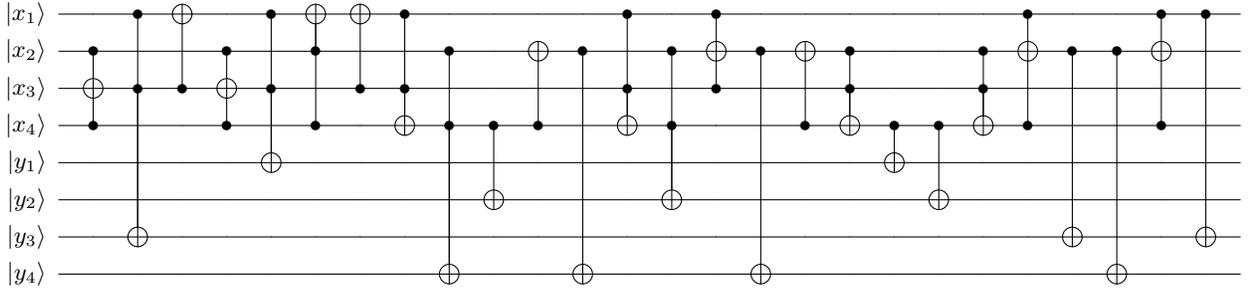
\begin{figure*}
\centerline{
\Qcircuit @C=1em @R=.7em {
\lstick{\ket{x_1}} &\qw &\ctrl{6} &\targ &\qw &\ctrl{4} &\targ &\targ &\ctrl{3} &\qw &\qw &\qw &\qw &\ctrl{3} &\qw &\ctrl{1} &\qw &\qw &\qw &\qw &\qw &\qw &\ctrl{1} &\qw &\qw &\ctrl{1} &\ctrl{6} &\qw \\
\lstick{\ket{x_2}} &\ctrl{1} &\qw &\qw &\ctrl{1} &\qw &\ctrl{-1} &\qw &\qw &\ctrl{6} &\qw &\targ &\ctrl{6} &\qw &\ctrl{4} &\targ &\ctrl{6} &\targ &\ctrl{2} &\qw &\qw &\ctrl{2} &\targ &\ctrl{5} &\ctrl{6} &\targ &\qw &\qw \\
\lstick{\ket{x_3}} &\targ &\ctrl{4} &\ctrl{-2} &\targ &\ctrl{2} &\qw &\ctrl{-2} &\ctrl{1} &\qw &\qw &\qw &\qw &\ctrl{1} &\qw &\ctrl{-1} &\qw &\qw &\ctrl{1} &\qw &\qw &\ctrl{1} &\qw &\qw &\qw &\qw &\qw &\qw \\
\lstick{\ket{x_4}} &\ctrl{-1} &\qw &\qw &\ctrl{-1} &\qw &\ctrl{-3} &\qw &\targ &\ctrl{4} &\ctrl{2} &\ctrl{-2} &\qw &\targ &\ctrl{2} &\qw &\qw &\ctrl{-2} &\targ &\ctrl{1} &\ctrl{2} &\targ &\ctrl{-2} &\qw &\qw &\ctrl{-2} &\qw &\qw \\
\lstick{\ket{y_1}} &\qw &\qw &\qw &\qw &\targ &\qw &\qw &\qw &\qw &\qw &\qw &\qw &\qw &\qw &\qw &\qw &\qw &\qw &\targ &\qw &\qw &\qw &\qw &\qw &\qw &\qw &\qw \\
\lstick{\ket{y_2}} &\qw &\qw &\qw &\qw &\qw &\qw &\qw &\qw &\qw &\targ &\qw &\qw &\qw &\targ &\qw &\qw &\qw &\qw &\qw &\targ &\qw &\qw &\qw &\qw &\qw &\qw &\qw \\
\lstick{\ket{y_3}} &\qw &\targ &\qw &\qw &\qw &\qw &\qw &\qw &\qw &\qw &\qw &\qw &\qw &\qw &\qw &\qw &\qw &\qw &\qw &\qw &\qw &\qw &\targ &\qw &\qw &\targ &\qw \\
\lstick{\ket{y_4}} &\qw &\qw &\qw &\qw &\qw &\qw &\qw &\qw &\targ &\qw &\qw &\targ &\qw &\qw &\qw &\targ &\qw &\qw &\qw &\qw &\qw &\qw &\qw &\targ &\qw &\qw &\qw
}
}
\caption{The quantum circuit for finding the inverse in $GF(2^4)$}
\label{FI}
\end{figure*}

We design a quantum circuit in Figure \ref{FI} based on the Equation (\ref{BF}), which does not require auxiliary qubits or the initial state of the output qubits in $\ket{0}^{\otimes 4}$. If the initial state of the output qubits are not in $\ket{0}^{\otimes 4}$, the results are equal to performing XOR operations with the initial state of the output qubits.
This process requires 14 Toffoli gates and 11 C-NOT gates. 

\begin{figure}[H]
\centerline{
\Qcircuit @C=1em @R=.7em {
\lstick{\ket{y_1}} &\targ &\ctrl{3} &\targ &\ctrl{1} &\targ &\qw &\qw &\targ &\gate{X} &\qw &\rstick{\ket{a_3}} \\
\lstick{\ket{y_2}} &\qw &\qw &\ctrl{-1} &\targ &\ctrl{-1} &\targ &\qw &\qw &\qw &\qw &\rstick{\ket{a_2}} \\
\lstick{\ket{y_3}} &\qw &\qw &\qw &\qw &\qw &\ctrl{-1} &\targ &\ctrl{-2} &\qw &\qw &\rstick{\ket{a_1}} \\
\lstick{\ket{y_4}} &\ctrl{-3} &\targ &\qw &\qw &\qw &\qw &\ctrl{-1} &\qw &\gate{X} &\qw &\rstick{\ket{a_0}}
}
}
\caption{The merged quantum circuit between the affine transformation and the mapping matrix}
\label{ATM}
\end{figure}
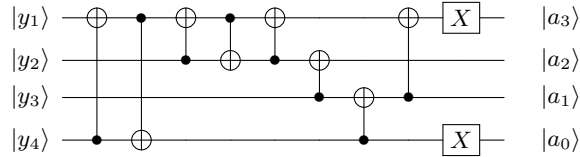

In addition, the affine transformation and the mapping matrix can be merged. The quantum circuit is shown in Figure \ref{ATM}, which only requires 8 C-NOT gates and 2 NOT gates. So the SubBytes in S-AES requires 14 Toffoli gates, 23 C-NOT gates and 2 NOT gates.

\section{Complexity analysis}
In AES-128, there are 256 SubBytes in 10-round encryption and 72 SubBytes in key expansion, where the first, second, third and sixth rounds sub-key need to be generated twice and elimination once. So performing the XOR operations with the known 32-bit string in key expansion requires 24 C-NOT gates. In addition, the operation $P$, its gate complexity at most $128 $ X gates, is performed four times. The number of times each operation needs to be performed and the gates required per operation are shown in Table \ref{OT}. In S-AES, there are 12 SubBytes. The number of times each operation needs to be performed and the gates required per operation are shown in Table \ref{SOT}. 
\begin{table}[H]
\begin{tabular}{|c|c|c|c|c|}
\hline
            & Toffoli & C-NOT & X gate & Times \\ \hline
IK &         &       & 128               & 1     \\ \hline
$P$           &         &       & $\leq$128    & 4     \\ \hline
ARK         &         & 128   &                   & 16    \\ \hline
SB       & 55      & 314   & 4                 & 256   \\ \hline
$SB^*$       & 55      & 322   & 4                 & 72    \\ \hline
MC          &         & 277   &                   & 60    \\ \hline
RC          &         &       &24                 &       \\ \hline
Others      &         & 96    &                   & 18    \\ \hline
\end{tabular}
\caption{AES-128: IK represents the initialization of key; $P$ represents the XOR operations between the plain-text and the initial key; ARK represents the AddRoundKey; SB represents the SubBytes in the round encryption; $SB^*$ represents the modified SubBytes in the key expansion; MC represents the MixColumns; RC represents the XOR operations with the known 32-bits string which needs 24 single qubit gates in the entire process(the gate used is different in each round); Others represent the remaining XOR operation in each key expansion.}
\label{OT}
\end{table}

\begin{table}[H]
\begin{tabular}{|c|c|c|c|c|}
\hline
            & Toffoli & C-NOT & X gate & Times \\ \hline
IK &         &       & 16                & 1     \\ \hline
$P$           &         &       & $\leq$16     & 2     \\ \hline
ARK         &         & 16    &                   & 2     \\ \hline
SB          & 14      & 23    & 2                 & 12    \\ \hline
MC          &         & 20    &                   & 2     \\ \hline
RC          &         &       & 3                 &       \\ \hline
Others      &         & 8     &                   & 2     \\ \hline
\end{tabular}
\caption{S-AES: SB represents the SubBytes; RC needs 3 single qubit gates in the entire process; Others represent the remaining XOR operation in each key expansion, and the other abbreviations are same as the Table \ref{OT}}
\label{SOT}
\end{table}

Table \ref{CCB} compares the resource usage between Langenberg et al.'s work and this work in quantum implementation of AES-128. Table \ref{SCCB} compares the resource usage between Almazrooie et al.'s work and this work in quantum implementation of S-AES. 
\begin{table}[H]
\begin{tabular}{|c|c|c|c|c|}
\hline
              & Toffoli & C-NOT  & X gate & Qubits \\ \hline
Langenberg et al. & 16940 & 107960 & 1507   & 880              \\ \hline
This work    & 18040  & 101174  &1976    & 656              \\ \hline
\end{tabular}
\caption{AES-128: The cost comparison between Langenberg et al.'s work and this paper}
\label{CCB}
\end{table}

\begin{table}[H]
\begin{tabular}{|c|c|c|c|c|}
\hline
                  & Toffoli & C-NOT & X gate & Qubits \\ \hline
Almazrooie et al. & 576     & 1080   & 26     & 72                \\ \hline
This work        & 168     & 364   & 75     & 48                \\ \hline
\end{tabular}
\caption{S-AES: The cost comparison between Almazrooie et al.'s work and this paper}
\label{SCCB}
\end{table}

A comparison in Table~\ref{CCB} shows that the gate complexity between Langenberg et al.'work and our work is basically the same, but 224 qubits are saved. Table \ref{SCCB} shows that the quantum circuits we proposed for S-AES can save $71\%$ Toffoli gates, over $66\%$ C-NOT gates, and one third of the qubits usage compared to Almazrooie et al.'s work.

\section{\label{sec5:level1}discussion \& conclusion}
In this work, we proposed an improved key expansion algorithm and modified the SubBytes circuit in Langenberg et al.'s work. The design for the key expansion algorithm was structured to utilize the lowest possible number of qubits, which cuts the number of qubits by 224 for AES-128. All of which advance the quantum implementation of AES.
We applied the same methods to S-AES and got a similar result. The reduction of the number of qubits makes S-AES algorithm possible to be implemented on quantum computers in the near future, while the cost of 48 qubits also allows classical computers to do the full amplitude quantum simulation.

At present, the quantum attack on AES algorithm is mainly based on Grover's algorithm, while the quantum implementation of AES algorithm, as an Oracle, is the most important part in Grover attack. We have constructed a simpler Oracle that will accelerate the process of quantum Grover attack.

After our work was done, we noted that a similar paper was published\cite{zou2020quantum}. There are also two main improvements compared with Langenberg et al.'s work. One improvement is in sub-key expansion. They solved the problem that the initial state of the output qubits in SubBytes is not in $\ket{0}^{\otimes 8}$ state by introducing an auxiliary qubit, while ours is making an affine transformation. The other improvement is proposing a new zig-zag method which can reduce 256 qubits in the encryption process compared with Grassl et al.s work. If the new zig-zag method was used in our work, the quantum resource usage can be further reduced. Table \ref{Zou} shows the number of times each operation needs to be performed and the gates required per operation using the new zig-zag method. The resource usage in using new zig-zag method is shown in Table \ref{newsource}. 
\begin{table}[H]
\begin{tabular}{|c|c|c|c|c|}
\hline
                           & Toffoli & C-NOT & X gate & Times \\ \hline
IK                         &         &       & 128               & 1     \\ \hline
$P$                          &         &       & $\leq$128   & 4     \\ \hline
ARK                        &         & 128   &                   & 10    \\ \hline
SB(r)                      & 55      & 314   & 4                 & 160   \\ \hline
SB(k)                      & 55      & 322   & 4                 & 40    \\ \hline
$SB^{-1}$ & 63      & 341   & 24                & 128   \\ \hline
MC                         &         & 277   &                   & 36    \\ \hline
RC                         &         &       & 16                &       \\ \hline
Others                     &         & 96    &                   & 10    \\ \hline
\end{tabular}
\caption{AES-128 with new zig-zag method: IK represents the initialization of key; $P$ represents the XOR operations between the plain-text and the initial key; ARK represents the AddRoundKey; SB(r) represents the SubBytes in the round encryption; SB(k) represents the SubBytes in the key expansion; $SB^{-1}$ represents the inverse process of SubBytes in \cite{zou2020quantum}. MC represents the MixColumns; RC represents the XOR operations with the known 32-bits string which needs 16 single qubit gates in the entire process(the gate complexity is different in each round); Others represent the remaining XOR operation in each key expansion.}
\label{Zou}
\end{table}

\begin{table}[H]
\begin{tabular}{|c|c|c|c|c|c|}
\hline
                & Toffili & C-NOT & X gate & Qubits & Auxiliary qubits \\ \hline
Zou et al. & 19788   & 128517   & 4528 & 512    & 0                \\ \hline
This work      & 19064   & 118980   & 4528 & 384    & 16               \\ \hline
\end{tabular}
\caption{AES-128: the quantum resource usage using new zig-zag method}
\label{newsource}
\end{table}
Table~\ref{newsource} shows that our work is slightly prior to Zou et al.'s work in gate complexity and reduces the qubits used by 128.

\textbf{Acknowledgements---}

We gratefully acknowledges support from the National Natural Science Foundation of China under Grants No. 11974205, and No. 11774197. The National Key Research and  Development Program of China (2017YFA0303700); The Key Research and  Development Program of Guangdong province (2018B030325002); Beijing Advanced Innovation Center for Future Chip (ICFC).
S.W. also acknowledges  the China Postdoctoral Science Foundation 2020M670172 and the National Natural Science Foundation of China under Grants No. 12005015.

\bibliographystyle{unsrt}
\bibliography{draft-arxiv}

%


\end{document}



\title{Supplementary for "Quantum Gradient Algorithm for General Polynomials"}

\date{\today}

\maketitle
\tableofcontents

\section{\label{sec2:level1} Gradient-based Algorithms}
Gradient algorithms, a first-order method, is widely applied in problems such as
\begin{eqnarray}
\max_{\bm{x}} f(\bm{x}) \quad \mbox{or} \quad  \min_{\bm{x}}  f (\bm{x}) , \quad \mbox{for all} \quad \bm{x} \in \mathbb{R}^d,
\end{eqnarray}
where we confine the $d$-dimension variable $\bm{x}$ to the real 
since the complex numbers can be processed with two independent real ones. Thus the problems can be solved iteratively with an iterative equation which is defined as
\begin{eqnarray}
\bm{x}^{t+1}=\bm{x}^t \pm \eta \bm{\nabla} f(\bm{x}^t),
\end{eqnarray}
where $\eta$ is dubbed as the learning rate, the superscript $^t$ is labelled as the $t$-th step of iteration and $-(+)$ corresponds to the minimum(maximum) problems, respectively.

Various algorithms, derived from the prototypical gradient method, is playing an important role in nowadays machine-learning technology. Those gradient-based algorithms include fist-order methods such as Vanilla, Stochastic, Mini-batch, Momentum, RMSprop and Adam methods, and the second-order methods such as Newton\cite{2019-Zhang-GforDP}. However, they are suffering the unmanageable consumption of either time or space for high-dimension parameter optimization since the typical gradient vector $\bm{\nabla}f(\bm{x})$ in step $t$ is usually approximated by a numerical differentiation methods as
\begin{eqnarray}
\frac{\partial _i f(\bm{x})}{\partial x_i} |_{\bm{x}^t}= \lim_{\delta \rightarrow 0}\frac{f( \cdots,  x^t_i +\frac{\delta}{2} ,\cdots)
-f(\cdots, x^t_i -\frac{\delta}{2} ,\cdots)} { \delta}.
\end{eqnarray}
This is known as symmetric difference quotient for discarding of second-order term and involves $2d$ times query of Oracle $O_{f}$ which act as $O_f(\bm{x})=f(\bm{x})$. 
So, when the variable dimension $d$ grows larger and larger, especially being in billions for nowadays machine-learning, extremely high cost of time and space is required and the method is thus out of efficiency.

\section{\label{sec3:level1} Framework of Quantum Gradient Descent Algorithm}
\subsection{\label{Binary_encoding}Binary coding quantum algorithm}
We review the framework of quantum gradient descent method proposed by Jordan, who encodes the variable in the binary from\cite{2005-jordan-Qgradient} 
\begin{eqnarray}
\bm{x}^t \rightarrow \ket{\bm{x}^t}=\ket{x^t_1}\otimes \ket{x^t_2} \otimes \cdots \otimes \ket{x^t_d} .
\end{eqnarray}
To assist the implementation of the gradient method, an Oracle $U_f$ which act as
\begin{eqnarray}\label{eqOf}
U_f: \quad U_f\ket{\bm{x}^t}=e^{i \frac{2\pi}{N_o}\frac{\eta}{2} f(\bm{x}^t)} \ket{\bm{x}^t}
\end{eqnarray}
is defined and could be realized by phase-kicking with ancilla qubits in state $\ket{\psi}=\frac{1}{\sqrt{2N_o/\eta}}\sum_{a=0}^{\frac{2N_o}{\eta} -1}e^{-i \frac{2\pi}{N_o} \frac{\eta}{2} a} \ket{a}$ as 
\begin{eqnarray}
 &&\ket{\bm{x}^t}\otimes \ket{\psi} 
 \xrightarrow{f}  \ket{f(\bm{x}^t)} \otimes \ket{\psi} \nonumber \\
 \xrightarrow{C^{-1}_{not}}&&\ket{f(\bm{x}^t)}\otimes \frac{1}{\sqrt{2 N_o/ \eta}}\sum_{a=0}^{2N_o/\eta -1} e^{i\frac{2\pi}{2N_o/\eta}a}\ket{a-f(\bm{x}^t)}   \nonumber \\
\xrightarrow{f^{-1}}&&e^{i\frac{2\pi}{N_o} \frac{\eta}{2}f({\bm{x}^t})}\ket{\bm{x}^t}\otimes \ket{\psi}
\end{eqnarray}
And also, many other type modular addition can be done by modified control-Not gates, that is, phase kicking is a flexible method which generally used to exponentiate binary numbers as phase value in quantum computation world. With the hand of this, the full algorithm is given conditionally on the ancillary qubits as 
\begin{eqnarray}
initial:\quad &&\ket{\bm{x}^t} \otimes \ket{0}   \nonumber \\
 \xrightarrow{Hardmard}\quad &&\sum_{\bm{\delta}}\ket{\bm{x}^t} \otimes \ket{\bm{\delta}}   \nonumber \\
 \xrightarrow{C_{not}}\quad &&\sum_{\bm{\delta}}\ket{\bm{x}^t+ \xi \bm{\delta}} \otimes \ket{\bm{\delta}}   \nonumber \\
 \xrightarrow{U_f}\quad &&\sum_{\bm{\delta}}e^{i\frac{2\pi}{N_0} \frac{\eta}{2} f(\bm{x}^t+\xi \bm{\delta})} \ket{\bm{x}^t+ \xi \bm{\delta}} \otimes \ket{\bm{\delta}}   \nonumber \\
 \xrightarrow{cnot^{-2}}\quad &&\sum_{\bm{\delta}}e^{i\frac{2\pi}{N_0} \frac{\eta}{2} f(\bm{x}^t+\xi \bm{\delta})} \ket{\bm{x}^t- \xi \bm{\delta}} \otimes \ket{\bm{\delta}}   \nonumber \\
\xrightarrow{U_f^{-1}}\quad &&\sum_{\bm{\delta}}e^{i\frac{2\pi}{N_0} \frac{\eta}{2} [f(\bm{x}^t+\xi \bm{\delta})-f(\bm{x}^t-\xi \bm{\delta})]} \ket{\bm{x}^t- \xi \bm{\delta}} \otimes \ket{\bm{\delta}}   \nonumber \\
\xrightarrow{cnot}\quad &&\sum_{\bm{\delta}}e^{i\frac{2\pi}{N_0} \eta \xi \bm{\delta} \cdot \bm{\nabla} f(\bm{x}^t)} \ket{\bm{x}^t} \ket{\bm{\delta}} \nonumber \\
=\quad &&\ket{\bm{x}^t} \ket{ \eta \xi \bm{\nabla}f(\bm{x}^t)} \nonumber \\
\xrightarrow{cnot}\quad &&\ket{\bm{x}^t+\eta \xi \bm{\nabla}f(\bm{x}^t)}  \ket{\eta \xi \bm{\nabla}f(\bm{x}^t)}
\end{eqnarray}
where $\eta$ scales the step size while $\xi$ should be small enough for efficiency approximation of the gradient and mitigating the effect from non-linear term.

Bit-consumption is exactly the same as the classical methods. To preserve a $d$-dimension real variable to an accuracy $2^{-\chi}$, $d\chi$ qubits are required. As for the gate complexity, $O(d+log_{2}d)$ quantum gates are required from quantum Fourier transformation, just maintaining a same level as the classical algorithm. However, query complexity is reduced to only 2 for calling oracle $O_f$ while $2d$ queries are needed in classical situation.

\subsection{\label{Amplitude_encoding} Amplitude coding quantum algorithm}
Variables are usually encoded as qubit-amplitude in quantum information processing for resources reduction and speed up. Rebentrost et al, absorbed it in their original quantum gradient descent work, which dealing with $2p$-order homogeneous polynomials\cite{2019-Rebentrost-Qgradient}. This optimization problem of homogeneous polynomials can be  expressed in tensor language as
\begin{eqnarray}
 f(\bm{x}^t)&&=\sum_{m_1+\cdots+m_d=2p} a_{m_1,\cdots,m_d}(x_1^t)^{m_1}\cdots(x_d^t)^{m_d} \nonumber \\
&&=\frac{1}{2} \underbrace{\bm{x}^t\otimes \cdots \otimes \bm{x}^t}_{p} A \underbrace{\bm{x}^t\otimes \cdots \otimes \bm{x}^t}_{p} \nonumber \\
&&=\frac{1}{2}\sum_{\alpha=1}^K \prod_{n=1}^{p} \bm{x}^t A_n^{\alpha} \bm{x}^t
\end{eqnarray}
where $a_{m_1,\cdots,m_d}$ is the coefficient of the polynomials and $\bm{x}^t=(x_1^t,x_2^t,\cdots,x_d^t)$ is the multi-dimension variable. By involving the coefficients in a matrix $A$, the objective function can be seen as the component of $A$ in a tensor of $p$ $d$-dimension variable spaces. In general, $A$ can be composed of a sum of tensor product of unitary matrices $A=\sum_{\alpha=1}^K \prod_{n=1}^p \otimes A_n^{\alpha}$, where $A_n^{\alpha}$ acts on the $n$-th $d$-dimension variable space. 

With above depiction, the gradient of $f(\bm{x}^t)$ can be explicitly determined by the current position $\bm{x}^t$
\begin{eqnarray}
\bm{\nabla} f(\bm{x})=\sum_{\alpha=1}^{K}\sum_{n_1=1}^{p} A_{n_1}^{\alpha} \bm{x}^t\prod_{n_2=1,n_2\ne n_1}^p \bm{x}^t A_{n_2}^{\alpha} \bm{x}^t
\end{eqnarray}

The corresponding quantum amplitude encoding expressed as
\begin{eqnarray}
\bm{x}^t=(x_1^t,x_2^t,\cdots,x_{d}^t)   
\rightarrow \ket{x}^t=\sum_{i=1}^{d}x_i^t\ket{i}.
\end{eqnarray}
here we must point out this encoding way implies the normalization constraint $||\bm{x}^t||=1$. That is, the variable is constrained on a $d$-dimension sphere. It benefits the case where the results need not to be expressed explicitly but only used for further analysis, such as expectation value of some certain observables. In this way,  the objective function as expectation value and gradient state can be rewritten as 
\begin{eqnarray}
f(\ket{\bm{x}^t})=\frac{1}{2}\sum_{\alpha=1}^K \prod_{n=1}^p \bra{\bm{x}^t} A_n^{\alpha} \ket{\bm{x}^t} 
\end{eqnarray}
\begin{eqnarray}
\ket{\nabla f(\ket{\bm{x^t}})}&&=\frac{1}{\mathcal{N}} \sum_{\alpha=1}^{K}\sum_{n_1=1}^{p} \prod_{\substack{n_2=1, \\ n_2\ne n_1}}^p \bra{\bm{x}^t} A_{n_2}^{\alpha} \ket{\bm{x}^t} A_{n_1}^{\alpha} \ket{\bm{x}^t}=\frac{1}{\mathcal{N}}\hat{D}_{\ket{\bm{x}^t}} \ket{\bm{x}^t}
\end{eqnarray}
The key to obtain the iterative state is to apply the operator $\hat{D}_{\ket{\vec{x}^t}}$ to the current state. Generally, it is a position-dependent operator which could be unitary or not. With the help of Hamiltonian simulation, qPCA and $HHL$-like methods at a cost of extra multi-copies of $\ket{\vec{x}^t}$, this operator application can be achieved.

However, since each iteration multi-copies state are used, the algorithms would scale exponentially in the number of iterations performed.On the other hand, amplitude encoding may perform better in resources consumption, as the memory consumption is $O(poly(p)^t\log(d))$, where $t$ is the number of iterations. In cases where a reasonable solution can be obtained with a limited iteration, it is acceptable for the memory consumption with $O(poly \log(d))$. As for the gates complexity or query complexity, we leave them in our following work and analyze them as comparison.

In a word, for this method, the resources consumption scales logarithmically with variable dimension $d$ and exponentially with iteration steps. Thus, it performs well with large $d$ and fast-convergence problems.

\section{Subroutines}
After inspecting the three stages in our algorithm in the main text,  four stubborn problems are left to four subroutines. Here, we will show the corresponding details explicitly, including \emph{Subroutine-1}, for initialization of the quantum register;\emph{Subroutine-2}, for applying the $HHL$-like method;\emph{Subroutine-3}, for implementing the variable-dependent $e^{-i\mathcal{D}_{\ket{\bm{X}^t}}}$ with quantum principle component analysis method and \emph{Subroutine-4}, for constructing the Non-Unitary operation $\mathcal{K}$. In this section, we drop the \emph{step}-subscript or -superscript such as $^t$ and $_t$, and the \emph{variable}-subscript $_{\ket{\vec{X}}}$ for convenience. 

\subsection{Subroutine-1: initialization}

\emph{Subroutine-1} serves as an operator, which drives the formatted variable register $v$ to the current variable state $\ket{\bm{X}}=\cos{\gamma}(\ket{0}+\sum_{i=1}^d x_i \ket{i})$. Typically, the concerned optimal problems are usually insensitive to the initial variables, so we can start with some product state, eg., $\ket{0}^{\otimes d+1}$. Thus, the complexity of this subroutine can be ignored in the first iteration step. As for the following iterations, output of the last iteration can play the role of this subroutine. 

Even in some cases we have to start with particular state $\ket{\psi}$, the recent-published result shows only  $\mathcal{O}(poly\log d)$  times of quantum manipulations are required with the help of 'bucket brigade' architecture of quantum random access memory(qRAM) per memory call\cite{2008-giovannetti-Qram,2012-Wu-RobustQRAM}.  An alternative way that based on Grover search can also generate arbitrary quantum state with suitably bounded amplitude with fidelity nearly to 1 by consuming $\mathcal{O}(poly\log d)$ qubits\cite{2006-soklakov-SPboundamp}. Hence a conservative estimate of \emph{Subroutine-1} is admitted within $\mathcal{O}(poly\log d)$ time complexity with error assumed as $\epsilon_{I}$ in this process.

\subsection{Subroutine-2: construction of $D$}
There are two quantum phase estimation modules and one double control operation in \emph{Subroutine-2}. Suppose we have easy-accessible controlled-unitary operation $C_{\mathcal{U_D}}=\ket{1}\bra{1}\otimes \sum_{j=0}^{2^{\chi}-1} \ket{j}\bra{j} e^{-i \frac{2\pi}{2^{\chi}}\mathcal{D}j}+\ket{0}\bra{0}\otimes \mathbb{I}$ (whose implementation will be specified in the next \emph{Subroutine-3}), non-unitary operation $\mathcal{D}$ can usually be constructed by a $HHL$-like method. 
After a single qubit rotation on register $d_1$ which functions as
\begin{eqnarray}
    \ket{0}_{d_1} \ket{0}_{d_2} \ket{0}_e \ket{\bm{X}}_v
    \xrightarrow{R^x_{d_1}(\eta)}&&(\cos{\eta}\ket{0}_{d_1}+\sin{\eta}\ket{1}_{d_1})\ket{0}_{d_2}\ket{0}_e \ket{\bm{X}}_v,
\end{eqnarray}
this method is sketched with the following derivation.

\emph{First}, the binary estimation of the eigenvalues of $\mathcal{D}$ can be resolved as $\lambda _k$(with corresponding eigenstate $\ket{k}$) to the eigenstate register $e$ by quantum phase estimation method. The Hadamard gates $H_e$, control unitary evolution  $C_{\mathcal{U_D}}$ and controlled quantum Fourier transformation on $e$, $C_{\mathcal{U}^{-1}_F}$(functioned at state $\ket{1}_{d_1}$) are employed here in sequence. We summarize this procedure as 
\begin{eqnarray}
\xrightarrow{H_e}&&\frac{1}{\sqrt{2^{\chi}}}(\cos{\eta}\ket{0}_{d_1}+\sin{\eta}\ket{1}_{d_1})\ket{0}_{d_2}\sum_{j=0}^{2^{\chi}-1}\ket{j}_e \ket{\bm{X}}_v\nonumber \\
\xrightarrow{C_{\mathcal{U_D}}}&&\frac{\cos{\eta}}{\sqrt{2^{\chi}}}\ket{0}_{d_1}\ket{0}_{d_2}\sum_{j=0}^{2^{\chi}-1}\ket{j}_e \ket{\bm{X}}_v
+\frac{\sin{\eta}}{\sqrt{2^{\chi}}}\ket{1}_{d_1}\ket{0}_{d_2}\sum_{k=1}^{d+1}\sum_{j=0}^{2^{\chi}-1}\ket{j}_{e} e^{-i\frac{2\pi}{2^\chi}\lambda_k j} \beta_k\ket{k}_v\nonumber \\
\xrightarrow{C_{\mathcal{U}^{-1}_F}}&&\frac{\cos{\eta}}{\sqrt{2^{\chi}}}\ket{0}_{d_1}\ket{0}_{d_2}\sum_{j=0}^{2^{\chi}-1}\ket{j}_e \ket{\bm{X}}_v +\frac{\sin{\eta}}{\sqrt{2^{\chi}}}\ket{1}_{d_1}\ket{0}_{d_2}\sum_{k=1}^{d+1}\ket{\lambda_k}_e \beta_k\ket{k}_v
\end{eqnarray}

\emph{Then}, a multi-control rotation depends on both register $d_1$ and $e$, with angle $\theta=\arccos{\lambda_k}$ is applied on register $d_2$.
\begin{eqnarray} {\label{eigenvalue_rotation}}
\xrightarrow{C_{R^x(\theta)}}&& \frac{\cos{\eta}}{\sqrt{2^{\chi}}}\ket{0}_{d_1}\ket{0}_{d_2}\sum_{j=0}^{2^{\chi}-1}\ket{j}_e \ket{\bm{X}}_v+\frac{\sin{\eta}}{\sqrt{2^{\chi}}}\ket{1}_{d_1}\sum_{k=1}^{d+1}(\lambda_k\ket{0}_{d_2}+\sqrt{1-\lambda_k^2}\ket{1}_{d_2})\ket{\lambda_k}_e \beta_k\ket{k}_v
\end{eqnarray}

\emph{Finally}, the inverse conditional phase estimation procedure is conducted and the register $e$ is uncoupled from the working system. 
\begin{eqnarray}
\xrightarrow{C_{\mathcal{U}_F}}&& \frac{\cos{\eta}}{\sqrt{2^{\chi}}}\ket{0}_{d_1}\ket{0}_{d_2}\sum_{j=0}^{2^{\chi}-1}\ket{j}_e \ket{\bm{X}}_v+\frac{\sin{\eta}}{\sqrt{2^{\chi}}}\ket{1}_{d_1}\sum_{k=1}^{d+1}(\lambda_k \ket{0}_{d_2}+\sqrt{1-\lambda^2_k}\ket{1}_{d_2})\sum_{j=0}^{2^{\chi}-1}\ket{j}_e e^{-i\frac{2\pi}{2^\chi}\lambda_k j} \ket{\bm{X}}_v\nonumber \\
\xrightarrow{H_e \cdot  C^{-1}_{\mathcal{U_D}}}&& \frac{\cos{\eta}}{\sqrt{2^{\chi}}}\ket{0}_{d_1}\ket{0}_{d_2} \ket{0}_e  \ket{\bm{X}}_v+\frac{\sin{\eta}}{\sqrt{2^{\chi}}}\ket{1}_{d_1}\sum_{k=1}^{d+1}(\lambda_k \ket{0}_{d_2}+\sqrt{1-\lambda^2_k}\ket{1}_{d_2})\ket{0}_e \beta_k\ket{k}_v
\end{eqnarray}

A following rotation $R^{x}_{d_1}(\eta)$(or its inverse $(R^{x}_{d_1}(\eta))^{-1}$) on register $d_1$ with post selection of $\ket{0}_{d_1}\ket{0}_{d_2}$ will reduce the result above to be effective $\frac{1}{\mathcal{N}}\left( \ket{\bm{X}}_v-\tan^2{\eta} \mathcal{D} \ket{\bm{X}}_v \right)$ (or $\frac{1}{\mathcal{N}}\left( \ket{\bm{X}}_v+\tan^2{\eta} \mathcal{D} \ket{\bm{X}}_v \right)$ ), that is, the corresponding update of gradient descent(or ascent). To be noticed that for the symmetry of this subroutine, all the operation dependency on register $d_1$ except $C_{R^x(\theta)}$ in eq.(\ref{eigenvalue_rotation}) can be omitted with the result unchanged. Moreover, it will make a more flexible realization.

When we talk about the complexity, you will find that $C_{R^x(\theta)}$ can be implemented within $\chi$ two-qubit control rotations. So, with the available $C_\mathcal{U_D}$(the complexity of whose implementation will be discussed in the next subroutine), remaining resources consuming of this subroutine comes from standard \emph{phase estimation} shown as,

\emph{\textbf{Memory utilization}} 
\begin{eqnarray}
\chi=n_p+ \ulcorner log(2+\frac{1}{2\delta_{P}})\urcorner
\end{eqnarray}
qubits are required for binary storing of $\{\lambda_k\}$ with accuracy $\epsilon_p=2^{-n_p}$ and lower bound successful probability $1-\delta_{P}$. 

\emph{\textbf{Time consumption}} 
\begin{eqnarray}
 \mathcal{O}(\chi^2)   \mbox{ gates}
\end{eqnarray}

\subsection{Subroutine-3: Quantum principle component analysis}
In this subroutine, we show that the variable dependent operator $\sum_{j=0}^{2^\chi -1} \ket{j}\bra{j} e^{-i \mathcal{D} \frac{2\pi}{2^\chi} j}$ can be implemented within error $\epsilon_{\mathcal{D}}$ by \emph{Quantum Principle Component Analysis(qPCA)} methods.

First of all, we denote 
\begin{eqnarray}
\mathcal{M}=\sum_{k=1}^p  P_k A P^{\dagger}_k=\sum_{\alpha=1}^{K}\sum_{n_1=1}^{p} (\bigotimes_{\substack{n_2=1, \\ n_2\ne n_1}}^p  A_{n_2}^{\alpha})\otimes A_{n_1}^{\alpha} 
\end{eqnarray}
and $\rho=\ket{\bm{X}} \bra{\bm{X}}$, where $P_k$ be the permutation between the last and $k$-th subsystem.

By  simple derivation of \emph{qPCA} process, one can find that $\tau$-time evolution of the variable dependent gradient operator $\mathcal{D}=Tr_{p-1}[\rho^p \mathcal{M}]$ can be approximated as
\begin{widetext}
    \begin{eqnarray}\label{long_time_evolution}
    &&\underbrace{Tr_{p-1}\left[e^{-i \mathcal{M} \frac{\tau}{m}} \rho ^{\otimes p-1} Tr_{p-1}\left[e^{-i \mathcal{M} \frac{\tau}{m}}  \rho ^{\otimes p-1}\cdots
    Tr_{p-1}[  e^{-i \mathcal{M} \frac{\tau}{m}} \rho ^{\otimes p} e^{i \mathcal{M} \frac{\tau}{m}}]\cdots e^{i \mathcal{M}\frac{\tau}{m}} \right] e^{i \mathcal{M}\frac{\tau}{m}} \right]}_{m - \mbox{Trace}} \nonumber \\
    &&=(e^{-i \mathcal{D} \frac{\tau}{m}})^m \rho (e^{i \mathcal{D} \frac{\tau}{m}})^m + \mathcal{O}(m||\mathcal{D}||_{max}^2 \frac{\tau^2}{m^2}) \nonumber \\
    &&= e^{-i \mathcal{D}\tau} \rho e^{i \mathcal{D} \tau}+ \mathcal{O}(\frac{p^2 ||A||_{max}^2\tau^2}{m})
    \end{eqnarray}
    \end{widetext}

In a further step, $e^{-i\mathcal{M} \frac{\tau}{m}}=\prod_{k=1}^p P_k e^{-i A \frac{\tau}{m}} P_k+ \mathcal{O}(\frac{p^2||A||^2\tau^2}{m^2})$(exactly the same error as in eq.(\ref{long_time_evolution}), thus can be merged into each other latter) and $e^{-i A \frac{\tau}{m}}$ can be easily constructed with only access to the coefficient matrix $A$, that is, the well known two oracles
\begin{description}
\item[Oracle 1] $O_1 \ket{j,k}\ket{0}=\ket{j,k}\ket{A_{jk}}$
\item[Oracle 2] $O_2 \ket{j,l}=\ket{j,k_A(j,l)}$
\end{description}
where $k_A(j,l)$ denotes the column index of $l$-th nonzero element in $A$'s $j$-th row. 

By the results of Childs' and Low's work\cite{2015-DBerry-HSimulation,2017-chuang-Hsimulation}, one can efficient implement the sparse Hamiltonian simulation $e^{-i A \frac{\tau}{m}}$ with error $\epsilon_{hs}$ within 
$\mathcal{O}(\frac{s||A||\tau}{m} + \frac{\log{1/\epsilon_{hs}}}{\log{\log{1/\epsilon_{hs}}}})$ times queries of $O_1$and $O_2$, combined with $\mathcal{O}(p\log{(d+1)}+ \chi' poly\log{\chi'})$ gates, where $2^{-\chi'}$ be the Binary accuracy of $A$'s element as specified in $O_1$ and $s$ be $A$'s sparsity. We will roughly review this method in next section and the details can be find in \cite{2015-DBerry-HSimulation,2017-chuang-Hsimulation}.
 
According to above analysis, $p\pi$ time evolution of $A$, i.e. $e^{-iAp \pi}$ is involved in the implementation of $e^{-i \mathcal{D} \pi}$, simultaneously, the \emph{qPCA}-caused error can be derived as
\begin{eqnarray}
\epsilon_{pca}= \frac{\pi^2 p^2||A||^2_{max}}{m}
\end{eqnarray}
supposing we have exact Hamiltonian simulation. That is, we can take $m=\mathcal{O}(\frac{\pi^2 p^2||A||^2_{max}}{\epsilon_{pca}})$ for bounding the error purely comes from \emph{qPCA} process to be $\epsilon_{pca}$, and this process require
\begin{eqnarray}
\mathcal{O}(sp\pi||A||_{max}+\frac{\log{1/\epsilon_{hs}}}{\log{\log{1/\epsilon_{hs}}}})
\end{eqnarray}
times queries of $O_1$ and $O_2$.

Besides, notice that the control evolution can be constructed as
\begin{eqnarray}
\sum_{j=0}^{2^\chi-1} \ket{j}\bra{j} e^{-i \mathcal{D} \frac{2\pi}{2^\chi} j}= \sum_{\{j_k=0,1\}} \ket{j_1}\bra{j_1} e^{-i \mathcal{D} \frac{2\pi}{2} j_1} \cdots \ket{j_k}\bra{j_k} e^{-i \mathcal{D} \frac{2\pi}{2^k} j_k} \cdots \ket{j_\chi}\bra{j_\chi} e^{-i \mathcal{D} \frac{2\pi}{2^\chi} j_\chi}
\end{eqnarray}   
that is, the full error(both from \emph{qPCA} and \emph{Hamiltonian simulation}) of control evolution $\sum_{j=0}^{2^\chi-1} \ket{j}\bra{j} e^{-i \mathcal{D} \frac{2\pi}{2^\chi} j}$ can be bounded as
\begin{eqnarray}
\epsilon_{\mathcal{D}}=\epsilon_{pca} + \epsilon_{hs}
\end{eqnarray}
with the help of 
\begin{eqnarray}
\mathcal{O}\left(\chi( sp\pi ||A||_{max}+\frac{ \log{1/\epsilon_{hs}}}{\log{\log{1/\epsilon_{hs}}}} )\right)
\end{eqnarray}
times query of $O_1$ and $O_2$(for \emph{hamiltonian simulation}), and
\begin{eqnarray}
\mathcal{O}(\frac{\chi \pi^2 p^3 ||A||^2_{max} \log{(1+d)}}{\epsilon_{pca}})
\end{eqnarray}
two qubits swap gates(for implementation of $P_k$), and 
\begin{eqnarray}
\mathcal{O}(\frac{\chi \pi^2 p^3 ||A||^2_{max} \log{(1+d)}}{\epsilon_{pca}})
\end{eqnarray}
copies of state $\rho$.

\begin{figure}[!h]
    \centerline{
    \begin{tikzpicture}[thick]
      \tikzstyle{operator} = [draw,fill=white,minimum size=1.5em] 
      \tikzstyle{operator2} = [draw,fill=white,minimum size=4em] 
      \tikzstyle{operator3} = [draw,fill=white,minimum size=5em] 
      \tikzstyle{phase} = [fill,shape=circle,minimum size=3pt,inner sep=0pt]
      \tikzstyle{surround} = [fill=blue!10,thick,draw=black,rounded corners=2mm]
      \tikzstyle{ellipsis} = [fill,shape=circle,minimum size=2pt,inner sep=0pt]
      \node at (0.2,0.3) (note) {(a)};
      \node at (1.2,0) (q1) {$\ket{\vec{X}}$};
      \node at (1.2,-1) (q2) {$\ket{0}$};
      \node at (1.7,0) (hy11) {$/$};
      \node at (1.7,-1) (hy22) {$/$};
      \node (end1) at (3.5,0) {} edge [-] (q1);
      \node (end2) at (3.5,-1) {} edge [-] (q2);
      \node[operator2] (op21) at (2.5,-0.5) {$O_{\widetilde{G}}$} ;
      \node at (3.75,-0.5) (eq1) {$=$};
      \node at (4.5,0) (q11) {$\ket{\vec{X}}$};
      \node at (4.5,-1) (q22) {$\ket{0}$};
      \node at (5,0) (hy1) {$/$};
      \node at (5,-1) (hy2) {$/$};
      \node (end1) at (9.5,0) {} edge [-] (q11);
      \node (end2) at (9.5,-1) {} edge [-] (q22);
      \node[operator] (op2) at (6,-1) {$H^{\otimes{\chi}}$};
      \node[phase] (phase11) at (7,-1) {};
      \node[operator] (op5) at (7,0) {$U_D$};
      \node[operator2] (op3) at (8.5,-0.5) {$QFT^{-1}$};
      \draw[-] (op5) -- (phase11);
      \node at (0.2,-2.2) (note) {(b)};
      \node at (1.2,-3.25) (q3) {$\ket{\vec{X}}$};
      \node (end3) at (3.25,-3.25) {} edge [-] (q3);
      \node at (1.6,-3.25) (hy111) {$/$};
      \node[operator2] (op22) at (2.37,-3.25) {$U_D$} ;
      \node at (3.4,-3.25) (eq1) {$=$};
      \node at (3.9,-2.5) (q33) {$I$};
      \node at (3.9,-3) (q3232) {$\ket{\vec{X}}$};
      \node at (3.9,-4) (q44) {$\ket{\vec{X}}$};
      \node (end3) at (6.4,-2.5) {} edge [-] (q33);
      \node (end32) at (6.4,-3) {} edge [-] (q3232);
      \node (end4) at (6.4,-4) {} edge [-] (q44); 
      \node at (4.35,-2.5) (hy33) {$/$};
      \node at (4.35,-3) (hy3232) {$/$};
      \node at (4.35,-4) (hy44) {$/$};
      \node[operator3] (op21) at (5.3,-3.25) {$U_{M_D}$} ;
      \node[ellipsis] (ee1) at (3.9,-3.35) {};
      \node[ellipsis] (ee2) at (3.9,-3.5) {};
      \node[ellipsis] (ee3) at (3.9,-3.65) {};  
      \node at (6.5,-3.25) (eq1) {$=$};
      \node at (7,-2.5) (q33) {$I$};
      \node at (7,-3) (q3232) {$\ket{\vec{X}}$};
      \node at (7,-4) (q44) {$\ket{\vec{X}}$};
      \node (end3) at (9.5,-2.5) {} edge [-] (q33);
      \node (end32) at (9.5,-3) {} edge [-] (q3232);
      \node (end4) at (9.5,-4) {} edge [-] (q44); 
      \node at (7.45,-2.5) (hy33) {$/$};
      \node at (7.45,-3) (hy3232) {$/$};
      \node at (7.45,-4) (hy44) {$/$};
      \node[operator3] (op21) at (8.4,-3.25) {$U_{QAQ^{\dagger}}$} ;
      \node[ellipsis] (ee1) at (7,-3.35) {};
      \node[ellipsis] (ee2) at (7,-3.5) {};
      \node[ellipsis] (ee3) at (7,-3.65) {}; 
       \node at (10,0.3) (note) {(c)};
      \node at (10.5,0) (q3) {};
      \node at (10.5,-1) (q4) {};
      \node at (10.7,-1) (hyC1) {$/$};
      \node (end3) at (12.5,0) {} edge [-] (q3);
      \node (end4) at (12.5,-1) {} edge [-] (q4);   
      \node[operator2] (op22) at (11.5,-0.5) {$U_A$} ;
      \node at (12.75,-0.5) (eq1) {$=$};
      \node at (13,0) (q33) {};
      \node at (13,-1) (q44) {};
      \node at (13.2,-1) (hyC2) {$/$};
      \node (end3) at (16.5,0) {} edge [-] (q33);
      \node (end4) at (16.5,-1) {} edge [-] (q44);   
      \node at (16.5,0) (q333) {};
      \node at (16.5,-1) (q444) {};
      \node (end3) at (18.5,0) {} edge [-] (q333);
      \node (end4) at (18.5,-1) {} edge [-] (q444);   
      \node[operator2] (op21) at (14,-0.5) {$\hat{U}_{\phi_1}$} ;
      \node[operator2] (op21) at (15.5,-0.5) {$\hat{U}_{\phi_2}$} ;
      \node[operator2] (op21) at (17.5,-0.5) {$\hat{U}_{\phi_2}$} ;
      \node[ellipsis] (ee1) at (16.4,-0.5) {};
      \node[ellipsis] (ee2) at (16.5,-0.5) {};
      \node[ellipsis] (ee3) at (16.6,-0.5) {};    
       \node at (10,-2.2) (note) {(d)};
       \node at (10.5,-2.75) (q1) {};
       \node at (10.5,-3.75) (q2) {};   
       \node at (10.7,-3.75) (hyD1) {$/$};
       \node (end1) at (12.5,-2.75) {} edge [-] (q1);
       \node (end2) at (12.5,-3.75) {} edge [-] (q2);
       \node[operator2] (op21) at (11.5,-3.25) {$\hat{U}_{\phi}$} ;
       \node at (12.75,-3.25) (eq1) {$=$};
       \node at (13,-2.75) (q11) {};
       \node at (13,-3.75) (q22) {};
       \node at (13.2,-3.75) (hyD2) {$/$};
       \node (end1) at (18.5,-2.75) {} edge [-] (q11);
       \node (end2) at (18.5,-3.75) {} edge [-] (q22);
       \node[operator] (op1) at (14,-2.75) {$e^{i\phi\sigma_z/2 }$};
       \node[operator] (op2) at (15,-2.75) {$H$};
       \node[phase] (phase11) at (15.75,-2.75) {};
       \node[operator] (op5) at (15.75,-3.75) {$\hat{W}$};
       \node[operator] (op3) at (16.5,-2.75) {$H$};
       \node[operator] (op4) at (17.5,-2.75) {$e^{i\phi\sigma_z/2 }$};
       \draw[-] (op5) -- (phase11);
    \end{tikzpicture} } 
    \caption{The structure to implement the quantum quantum phase estimation module in $subroutine 2$.(a) is the standard quantum phase estimation, where $\mathcal{U_{D}}=e^{-i \hat{D} \tau}$ can be simulated in (b) with the $subroutine 3$, controlled qCPA and $\mathcal{M}$ simulation.(c) and (d) are the Hamiltonian simulation with quantum signal processing method.}   
    \label{oracle gradient} 
  \end{figure}

Once the process may fall short of our expectation, there are still other algorithm aiming at Hamiltonian simulation.  Advances in algorithms provide an alternative way to implement this simulation. For $A$ with different restrictive assumptions, several previous work has pave a way to handle it. These algorithms can be used to the problems with $k$-local interactions assumption, unitary model assumption and sparse matrix assumption, including Trotter-Suzuki decompositions, Taylor series expansion, quantum walk\cite{2010-Achilds-HSimulation} and quantum signal processing method\cite{2017-chuang-Hsimulation}.

\subsection{Subroutine-4: construction of $\mathcal{K}$}
Recall that since there is a projection constraint $||\bm{x}||=1$ in Lloyd's work\cite{2019_Rebentrost_Qgradient}, all variable lied in a high-dimension sphere and the iteration stopped at state $\ket{\bm{x}}\propto \mathcal{D}\ket{\bm{x}}$ since the projected gradient vanished.

However, as the projection constrained moved out in our work by the dressed amplitude encoding $\bm{x}\rightarrow \ket{\bm{X}}=\cos{\gamma}(\ket{0}+ \tan{\gamma} \sum_{i=1}^d \ket{i})$,  quantum state $\ket{\bm{X}^f}$ which satisfy $\ket{\bm{X}^f} \propto \ket{\bm{X}^f} - \mathcal{K} \mathcal{D} \ket{\bm{X}^f}$ be the new optimized result, instead of state $\ket{\bm{X}^{trap}}$ which satisfy $\mathcal{D}\ket{\bm{X}^{trap}} \propto \ket{\bm{X}^{trap}}$, where $\mathcal{K}=diag(0,1,1,\cdots,1)$. 

Non-unitary $\mathcal{K}$ can be well implemented with the help of one single ancilla qubit as
\begin{flalign}\label{post_selection}
\mathcal{K}\ket{\psi} \propto \ket{0}\bra{0}_a H_a C_E H_a \ket{0}_a\ket{\psi}
\end{flalign}
where $H_a$ be Hadamard and $C_E =\ket{0}\bra{0}_a \otimes \mathbb{I}+ \ket{1}\bra{1}_a \otimes diag(-1,1,1,\cdots,1)$(or in another view, $C_E=C^0_{Z_a}$, a control-Z gate on ancilla functioned when the principle system in state $\ket{0}$). The success probability of post selection on ancilla qubit state $\ket{0}$ equal to the weight of non-$\ket{0}$ conponent in $\ket{\psi}$, which match the applied condition of this optimal method, that is, the object function being fast converged (that is, with finite gradient).

Notice that $C_E=C^0_{Z_a}=X^{\otimes \log{(d+1)}} C^1_{Z_a} X^{\otimes \log{(d+1)}}$, where $C^1_{Z_a}$ can be easily implemented with $\mathcal{O}(\log{(d+1)})$ \emph{Toffoli} gates and $\mathcal{O}(\log{(d+1)})$ extra qubits which set in $\ket{0}$(or in another way, this can be done with $\mathcal{O}(\log^2{(d+1)})$ \emph{Toffoli} gates without ancillary).

\section{Hamiltonian simulation by quantum signal processing(QSP)}

In this section, the method of quantum signal processing to complete the evolution $e^{-iHt}$ of $d$-dimension Hamiltonian $H$( where $A$ play the role of $H$ in\emph{subroutine-3}) is introduced. This method is based on the sparse matrix assumption, which involves two oracles.
\begin{description}
    \item[Oracle 1] $O_1 \ket{j,k}\ket{0}=\ket{j,k}\ket{H_{jk}}$
    \item[Oracle 2] $O_2 \ket{j,l}=\ket{j,k_H(j,l)}$
\end{description}
where polynomial coefficients oracle $O_1$ operates on $\chi'+2\log{(d)}$ qubits with $i,j= 1, \cdots,  d$, to hold an accuracy of $H$'s elements to $2^{-\chi'}$. The function $k_H(j,l)$ output the column index of $l$-th nonzero element in $j$-th row of $H$. Sparse input oracle $O_2$ operates on $2\log{d}$ qubits and $l=1, \cdots, s$, where $s$ just be the sparsity of $H$.

The result shows one can efficient implement the sparse-$s$ Hamiltonian simulation $e^{-iHt}$ with error $\epsilon_{hs}$ within
\begin{description}
    \item[Query complexity] $\mathcal{O}(s||H||_{max}t + \frac{\log{1/\epsilon_{hs}}}{\log{\log{1/\epsilon_{hs}}}})$
    \item[Gate complexity] $\mathcal{O}(\log{d}+ \chi' poly\log{\chi'})$
\end{description}

In the following, we will sketch the outline to implement this method to Hamiltonian simulation. Fig.\ref{oracle gradient} (c) and (d) give the basic idea. Both circuits consist of two input, one ancillary qubit  and a functional workspace.  The series of $\hat{U}_{\phi}$ build up our simulation circuit, where $\hat{U}_{\phi}$ consists of the Hadamard gates, $z$-rotations and controlled-$\hat{W}$ operation. when the system input is on the eigenvectors $\ket{\lambda}$, i.e. $\hat{U}_{\phi}\ket{\lambda}=e^{\theta_{\lambda}}\ket{\lambda}$, it can be reduced to the single-qubit rotation $R_{\phi}(\theta_{\lambda})$ on the ancilla, where $R_{\phi}(\theta_{\lambda})=e^{-i(\theta_{\lambda}/2)(\sigma_x cos(\phi)+\sigma_y sin(\phi))}$.

A general $e^{ih(\theta_{\lambda})}$ can be approximated with $R_{\phi_1}(\theta_{\lambda})R_{\phi_2}(\theta_{\lambda})...R_{\phi_n}(\theta_{\lambda})$ via optimizing the parameters $\phi_1,\phi_2,...,\phi_n$ in the sequence of the $\hat{U}_{A}$ (shown in  the Fig.\ref{oracle gradient}(c)), which builds up a transformation on the functional workspace
\begin{eqnarray}
 W&=&\sum_{\lambda}e^{i\theta_{\lambda}}\ket{\lambda}\bra{\lambda}\nonumber \\
 \rightarrow V&=&\sum_{\lambda}e^{ih(\theta_{\lambda})}\ket{\lambda}\bra{\lambda}
\end{eqnarray}
When $h(\theta_{\lambda})=-\tau sin(\theta_{\lambda})$, the simulation circuit can be used in the Hamiltonian simulation via quantum walk. 

We will given a brief illustration about how to adopt above in the quantum walk. Given a $s$-sparse Hamiltonian $H$ acting on $d$ dimension Hilbert space. First of all, an ancillary qubit with the state $\ket{0}$ is appended, expanding the space from $d$ to $2d$, then the entire Hilbert space is duplicated(thus $\mathbb{C}^{2d} \otimes \mathbb{C}^{2d}$). The whole process can be done with the isometry $T$
\begin{eqnarray}
T=\sum_{j=0}^{d-1} \sum_{b \in \{0,1\}}  (\ket{j} \bra{j} \otimes \ket{b} \bra{b}) \otimes \ket{\phi_{j,b}}
\end{eqnarray}
where $\ket{\phi_{j,0}}$ and $\ket{\phi_{j,1}}$ are defined as 
\begin{eqnarray}
&&\ket{\phi_{j,0}}=\frac{1}{\sqrt{s}}\sum_{l \in S_j} \left( \ket{l}  \sqrt{\frac{H_{i,l}^{\*}}{X}}
\ket{0}+\sqrt{1-\frac{|H_{i,l}^*|}{X}}\ket{1} \right) \nonumber \\
&&\ket{\phi_{j,1}}=\ket{0} \ket{1} 
\end{eqnarray}
with $X\geq ||H||_{max}$ and $S_j$ be the set of indices of nonzero elements in column $j$ of $H$. This is a controlled state preparation step, performing on the input $\ket{j}\ket{b}$, to creat $\ket{\phi_{j,b}}$($b=0,1$). In the step,  one query to the oracle $O_1$ and $O_2$ and additional $(\chi'ploylog(\chi'))$ primitive gates are required.
After this stage, the unitary operator of the quantum walk are applied with
\begin{eqnarray}
U=i S (2TT^{\dagger}- \mathbb{I})
\end{eqnarray}
$S$ is swap operation, acting on the former and latter two register as $S \ket{j_1} \ket{b_1} \ket{j_2} \ket{b_2}=\ket{j_2} \ket{b_2} \ket{j_1} \ket{b_1}$ where $j_1,j_2 \in [d]$ and $b_1,b_2 \in \{0,1\}$.  As $U$ corresponds to reflection about $TT^{\dagger}$ followed by $S$, swapping ($2+2\log{d}$)-qubit registers, its query and gate complexities are identical to $T$ up to constant factors. 

Suppose that $\ket{\lambda}$ be $H$'s eigenstates as $H\ket{\lambda}=\lambda\ket{\lambda}$, The quantum walk operator $U$ and its eigenvalues $\mu_{\pm}$ thus satisfy the following rules
\begin{eqnarray}
&&U\ket{\mu_{\pm}}=\mu_{\pm}\ket{\mu_{\pm}}\nonumber \\
&&\ket{\mu_{\pm}}=(1+i\mu_{\pm}S)T\ket{\lambda} , \mu_{\pm}=\pm e^{\pm i \arcsin{(\lambda /X s)}}
\end{eqnarray}
Since 
\begin{eqnarray}
  || T \frac{1-iS}{\sqrt{2}} (iU)^{\tau}\frac{1+iS}{\sqrt{2}}T^{\dagger} -e^{i\tau sin^{-1}(H/||abs(H)||)}||<(||H||/||abs(H)||)^2
\end{eqnarray} 
If $\lambda /X s$ is small enough, applying $\tau=s||H||_{max}t$ steps of the discrete-time quantum walk $U$, we could simulate $t$ time Hamiltonian simulation since\cite{2010-Achilds-HSimulation}.
Therefore, finally, the inverse state preparation $T^{\dagger}$ is performed. For a successful simulation, the output should lie in the original space, and the ancillary qubit should be returned to the state $\ket{0}$. 

The hint is, the nonlinear of the phase factor $arcsin(\lambda/X s)$ making the simulation deviate from the desired value. The transformation via quantum signal processing from $W$ to $V$ is applied here, when $h(\theta_{\lambda})=-\tau sin(\theta_{\lambda})$. It modifies the simulation circuit to a controlled one(the simulation circuit is just the $W$ in above protocol).

\section{Success probability}
By simple derivation, the full output state in circuits in main text before $R_x(\eta)$ acted can be shown as
\begin{eqnarray}
\cos{\eta}\ket{0}_k \ket{0}_{d_1}\ket{0}_{d_2} \ket{0}_e \ket{\bm{X}}+ \sin{\eta}\ket{0}_k\ket{1}_{d_1}\ket{0}_{d_2}\ket{0}_e \mathcal{K}\mathcal{D}\ket{\bm{X}} \nonumber \\
+\sin{\eta} \ket{0}_k\ket{1}_{d_1}\ket{1}_{d_2}\ket{0}_e \mathcal{D}^{\perp}\ket{\bm{X}} +\sin{\eta} \ket{1}_k\ket{1}_{d_1}\ket{0}_{d_2}\ket{0}_e(\mathbb{I}-\mathcal{K})\mathcal{D}\ket{\bm{X}}
\end{eqnarray}
where $\mathcal{D}^{\perp}\ket{\bm{X}}$ denotes the rubbish state $\sum_k \sqrt{1-\lambda^2_k} \beta_k \ket{k}$.  
Obviously, when $R_x(\eta)$ applied and post selected on state $\ket{0}_k\ket{0}_{d_1}\ket{0}_{d_2}$, only the
\begin{eqnarray}
\cos^2{\eta}\ket{\bm{X}}-\sin^2{\eta} \mathcal{K} \mathcal{D} \ket{X}
\end{eqnarray}
preserved, with probability
\begin{eqnarray} \label{succ_probability}
P_{succ}&&=\cos^4{\eta}+\sin^4{\eta}|\mathcal{K} \mathcal{D}\ket{\bm{X}}|^2-\sin^2{\eta} \cos^2{\eta} (\bra{\bm{X}}\mathcal{K} \mathcal{D}\ket{\bm{X}}+\bra{\bm{X}}\mathcal{D} \mathcal{K}\ket{\bm{X}}) \nonumber \\
&&\geq \cos^4{\eta} + (\sin^4{\eta}-2\sin^2{\eta}\cos^2{\eta}) |\mathcal{K} \mathcal{D}\ket{\bm{X}}|^2
\end{eqnarray}

Typically, $\tan^2{\eta}$ as the learning rate usually be set a less than one value. That is, $\eta\in[0,\frac{\pi}{4}]$ and the second tern in eq.(\ref{succ_probability}) always be negative. Thus in a further step
\begin{eqnarray}
P_{succ} \geq (\cos^2{\eta} -\sin^2{\eta})^2.
\end{eqnarray}
This means, in principle, a larger learning rate $\tan^2{\eta}$ will cause a lower success probability $P_{succ}$. This is reasonable and does not give an upset result once we notice that it is a highly non-linear relationship between $\tan^2{\eta}$ and $P_{succ}$. In fact, even we set a relative large enough learning rate $\tan^2{\eta}=\frac{1}{3}$, there still be $P_{succ} \geq \frac{1}{4}$.

\section{Algorithm Simulation}
Based on the theoretical protocol, the simulation program consists of an iteration of three stages as depicted in the article. The program assumed that the Hamiltonian simulation, i.e. $e^{iDt}$ can be simulated directly in the machine. Therefore, two cases are simulated in the frame of our protocol.

The polynomial optimizations include both maximum and minimum problems, which state as 
\begin{eqnarray}
  max \quad f_1 &&=\frac{1}{2} (1,x) ^{\otimes 2}
  \begin{pmatrix}
  7/2 & 0 \\
  0 & -9/2
  \end{pmatrix}^{\otimes 2}
   (1,x)^{T\otimes 2} \nonumber \\
  min \quad f_2 &&=\frac{1}{2} (1,x_1,x_2) ^{\otimes 2}\left[
  \begin{pmatrix}
  1 & 0 & 0\\
  0 & 1 & 0\\
  0 & 0 & 1
  \end{pmatrix}^{\otimes 2} + 
  \begin{pmatrix}
  0 & 0 & 1\\
  0 & 0 & 0\\
  1 & 0 & 0
  \end{pmatrix}\otimes 
  \begin{pmatrix}
  0 & 0 & 0\\
  0 & 0 & 1\\
  0 & 1 & 0
  \end{pmatrix}\right]
  (1,x_1,x_2)^{T \otimes 2} 
\end{eqnarray}

According to the protocol, an arbitrary initial input $(1,\bm{x'})$ can be represented with DAE by a quantum state $\ket{\bm{X}}=\cos{\gamma}(\ket{0}+\sum_i x_i \ket{i})$. All projects are based on the circuit model, which consists of a set of elementary quantum gates, i.e. arbitrary single-qubit rotations and two-qubit controlled-unitary operators. 

Initialization was implemented by the rotations on a qubit or a qutrit, as measurement was realized by simulating tomography on the variable register $e$, after projecting entire system into the target subspace. As for the intermediate procedures, the gates are implemented as the algorithm circuit with the assumption to get $e^{iDt}$. 

Remarkably, the standard phase estimation module approximates the eigenvalue of the target unitary with a range of $(0,1)$. However, this module should be modified since when constructing the $D$, the gradient operator produces both the positive and negative component. The periodic property is utilized in our implementation and the eigenvalues in the gradient operator are normalized in a range of $(-\frac{1}{2},\frac{1}{2})$. Therefore, the output of this module is divided into two parts, results in $(0,\frac{1}{2})$ is directly readout while others in $(\frac{1}{2},1)$ should be resolved into the rangle of $(-\frac{1}{2},0)$ by the periodic condition. 

In the program, the simulation includes $15$ ancillary qubit in a conservative way, with $1$ for $k$, $1$ for $d1$, $1$ for $d2$ and $12$ for $e$, whose resolution has been discussed in article. The rest qubits are used to encode the variables. During the simulation, $\bm{x}=(1,7 \pm 3)$ are considered as the two random initial guess for $f_1$, and $\bm{x}=(1, \pm 5, \pm 5)$ are chosen as four different initial variables for $f_2$. And $0.05$($0.1$) is chosen as the learning rate for $f1$ ($f_2$).

$3$ types errors are investigated, including the initial error $\epsilon_I$, the operation error $\epsilon_{\mathcal{D}}$ and the phase estimation error $\epsilon_{ph}$. $\epsilon_I$ comes from the imperfection of initialization which cannot arrive at the target input. we simulate this situation repeatably for 20 times, by introducing a perturbation with uniform random distribution whose amplitude is $5\%$ to the input state. $\epsilon_{\mathcal{D}}$ is from the uncertainty of the $e^{iDt}$, as $e^{iDt}$ cannot be perfectly generated. We introduced the noise into the operator, with $1\%$ and $2\%$ to the amplitude of $D$ for two cases, and simulated them for 15 times. $\epsilon_{ph}$ is the truncation error which originated from the size of eigenvalue register $e$. To simulate this situation, we choose different sizes of the $e$ for $5$,$7$,$9$,$11$. On the other side, we investigate the effects on gradient value of different sizes of $e$.

\bibliographystyle{unsrt}
\bibliography{draft-arxiv}